# A novel self-locked energy absorbing system

Yuli Chen[a,]*, Chuan Qiao[a], Shougen Zhao[a], Cairu Zhen[a], Bin Liu[b]

[a] Institute of Solid Mechanics, Beihang University (BUAA), Beijing 100191, P.R.China

[b] AML, CNMM, Department of Engineering Mechanics, Tsinghua University, Beijing 100084, China

* To whom correspondence may be addressed: yulichen@buaa.edu.cn

**Abstract**

Metallic thin-walled round tubes are widely used as energy absorption elements. However, lateral splash of the round tubes under impact loadings reduces the energy absorption efficiency and may cause secondary damages. Therefore, it is necessary to assemble and fasten round tubes together by boundary constraints and/or fasteners between tubes, which increases the time and labor cost and affects the mechanical performance of round tubes. In an effort to break through this limitation, a novel self-locked energy-absorbing system has been proposed in this paper. The proposed system is made up of thin-walled tubes with dumbbell-shaped cross section, which are specially designed to interlock with each other and thus provide lateral constraint under impact loadings. Both finite element simulations and impact experiment demonstrated that without boundary constraints or fasteners between tubes, the proposed self-locked energy-absorbing system can still effectively attenuate impact loads while the round tube systems fail to carry load due to the lateral splashing of tubes. Furthermore, the optimal geometric design for a single dumbbell-shaped tube and the optimal stacking arrangement for the system are discussed, and a general guideline on the structural design of the proposed self-locked energy absorbing system is provided.

# 1 Introduction

Convenient and effective protections against collision, impact and blast loadings can greatly save lives and properties from car accidents and explosions (Chen et al., 2009; Fleck and Deshpande, 2004; Hanssen et al., 2006; Mills et al., 2009; Neves et al., 2010; Wasiowych et al., 1996; Yang et al., 2003). So it is of great importance to design stationary and temporary impact-resistant energy absorbing systems, which have attracted attentions of many researchers (Abramowicz, 2003; Alexander, 1960; Dharmasena et al., 2008; Evans et al., 2001; Ezra, 1972; Gibson and Ashby, 1997; Johnson and Mamalis, 1978; Jones, 2011; Jones and Wierzbicki, 2010; Langseth and Hopperstad, 1996; Lu and Yu, 2003; Qiu and Yu, 2011; Rawlings, 1974; Shukla et al., 2010). Many energy-absorbing systems are composed of light-weight structures, such as metallic foam(Gong et al., 2005; Tan et al., 2005), honeycomb(Fazekas et al., 2002; Hu et al., 2013), lattice(Fleck and Qiu, 2007; Zhang et al., 2008), sandwich plate (Xue and Hutchinson, 2003, 2004) and shell structures(Gupta and Gupta, 2014). The basic principle of these systems is to dissipate kinetic energy by plastic deformation of structures (Alghamdi, 2001; Lu and Yu, 2003; Olabi et al., 2007). Well-designed energy-absorbing system attenuate the impact effect in a designed way: not only a certain amount of impact energy can be absorbed, but also the impact force and deceleration can be controlled within a desired range(Carney III, 2010; Yu, 1986).

Metallic thin-walled tube is one of the most widely used energy absorbing structural elements in civil and defense engineering for its absorption characteristics and practical advantages, such as stable and long deformation process, small initial impact force, high specific energy dissipation, easy manufacturability, and low cost. Therefore, many existing impact energy absorbing systems are comprised of multi-row round tubes (Carney III, 2010; Shim and Stronge, 1986).

Many researchers have studied energy dissipation characteristics of tubular energy absorbers under axial loadings (Al Galib and Limam, 2004; Avalle and Belingardi, 1997; Chen and Wierzbicki, 2001; Guillow et al., 2001; Oshiro and Alves, 2007). An early investigation on the axial compression of a cylindrical shell was conducted by Alexander (1960). Based on the rigid perfect-plastic material and axisymmetric folding assumptions, the collapse load was expressed in terms of thickness and diameter. The axisymmetric folding mode is also called the concertina mode, and different failure mechanisms of this mode have been proposed by many authors (Abramowicz and Jones, 1986; Grzebieta, 1990; Wierzbicki et al., 1992). As summarized by Andrews et al. (1983), the folding mode can be categorized according to the wall thickness to diameter ratio $t/D$ and the length to diameter ratio $L/D$. Concertina folds usually occur in relatively thick-walled tubes and diamond folds are common in thin-walled tubes. Transition from concertina to diamond mode was also observed. Other axial failure modes, such as the expansion (Shakeri et al., 2007; Yang et al., 2010), inversion (Al-Hassani et al., 1972; Reid and Harrigan, 1998; Rosa et al., 2003) and splitting (Reddy and Reid, 1986; Stronge et al., 1983) of round tubes were also studied.

An early investigation on the lateral dynamic plastic flow buckling of a cylindrical metallic shell was carried out by Abrahamson and Goodier (1962), and the effect of tube wall thickness was considered in the lateral impact experiment of a single round tube. Later, the effect of elasticity as well as the strain-rate reversal was analyzed by Lindberg (1965), Lindberg and Kennedy (1975), Lindberg and Florence (1987) and Jones and Okawa (1976). Symonds (1965) discussed the impact of elastic waves on dynamic response of structures and indicated that the ultimate deformation mode is influenced by those waves. Through further research on the role of elastic waves and revision of the structural shock theory, a theoretical analysis performed by Reid and Bell (1984) revealed the significance of elastic waves on the deforming pattern and energy dissipation. Based upon these findings, Choi et al. (1986) investigated the strain-rate effect in steel tube clusters and indicated that neglect of strain rate would lead to underestimation of energy absorbing capacity. Duffey et al. (1990) and Florence et al. (1991) analyzed the effect of impact velocity variance on the buckling of shells. The lateral compression of a tube between rigid plates was analyzed by DeRuntz and Hodge (1963) using a rigid perfectly plastic model with 4 plastic hinges, and Redwood (1964) included the strain hardening effect in DeRuntz and Hodge's model. A different collapse mode with 6 plastic hinges was proposed by Burton and Craig (1963). The effect hardening was further studied by Reid and Reddy (1978) using plastic arc instead of plastic hinge.

Recently, researchers have been working on the improvement of the round tube system and proposed many new energy-absorbing devices based on round tubes. Shrive et al. (1984)

introduced a cylindrical system comprised of two concentric rings and a series of smaller tubes welded in between. The rings did not completely flattened under the impact load and the maximum opposing forces agrees with results from quasi-static tests. Reddy and Reid (1979) investigated the effect of side constraints for closed systems. Nested circular and non-circular tubes with indenters and constraints were examined by Morris et al. (2006) and Morris et al. (2007). Braided composite tubes were studied by Harte et al. (2000) and Xiao et al. (2009). Sandwich tubes were studied by Fan et al. (2011), Shen et al. (2015) and Baroutaji and Olabi (2014), and Fan et al. (2011) reported three major deformation patterns. Foam-filled tubes were studied by Niknejad et al. (2012) and compared with empty specimens. Besides, different geometries were also discussed. Corrugated tubes were investigated by Abdewi et al. (2008) and Eyvazian et al. (2012). Oblong tube were studied by Olabi et al. (2008) and Baroutaji et al. (2014) which demonstrated an almost ideal response when compressed between flat plates.

The energy absorbing systems composed of metallic thin-walled tubes now play an important role in impact resistance and protection fields. However, lateral splash of tubes under impact loadings reduces the energy absorption efficiency and may cause secondary damages. Therefore, it is necessary to assemble and fasten metallic thin-walled round tubes together (Yu, 1986). The conventional constraint methods can be classified into two types: (1) to apply high-strength steel walls to the boundary of tube arrays to restrict their lateral expansion (Knoell and Wilson, 1978), and (2) to fasten tubes by welding, rivets and/or bolts to constrain their relative motions (Carney III, 1987). Although round-tube energy-absorbing systems with both types of constraints has proved their reliability in practice, there are still many limitations which can be categorized into the following three aspects:

(1) Positioning structures, such as drilling, welding and other fasteners on the tube walls, affect the energy-absorbing characteristics of the tubes.

(2) The complicated assembly and constraints of round tube system increases the installation cost of time and labor.

(3) Round-tube energy-absorbing systems lack modifiability to respond quickly to the need of impact protection in emergencies since the scale and energy absorption capability of the system are determined by the assembling before application.

In order to overcome the limitations in practical applications of current round-tube energy-absorbing systems, this paper introduces a novel self-locked energy-absorbing system which is made up of the newly-designed dumbbell-shaped tubes, and suggests the optimal structural design of the system by studying its energy absorption characteristics. The paper is structured as follows. The self-locking design of the proposed system is introduced in Section 2. Section 3 presents the comparison between the responses of the self-locked system and round tube systems under impact loading, and Section 4 validates the self-locking effect of the proposed system by impact experiment. The optimal geometric design for a single tube and the optimal structural design for the system are discussed in Section 5. Conclusion and prospective are given in Section 6.

## 2 Self-locking design for energy absorbing structure

In order to achieve self-locking effect, a thin-walled tube with dumbbell-shaped cross section is proposed here as shown in Figs. 1(a) and 1(b). The dumbbell-shaped tube is comprised of two open cylindrical shells and two parallel flat plates. The cylindrical shells absorb impact energy through plastic deformation while the flat plates connect the cylindrical shells and maintain the geometry of tube. The geometry of the dumbbell-shaped tube can be defined by 5 parameters as in Fig. 1(b): the axial length $L$ of the dumbbell-shaped tube, the outer diameter $D$ of the cylindrical shell, the wall thickness $t$, the distance $A$ between two cylindrical shell axes and the spacing $P$ between two parallel flat plates including wall thickness. For brevity, in the following text, the 5 parameters will be referred to by their short names, i.e. the length $L$, the diameter $D$, the thickness $t$, the axis distance $A$ and the plate spacing $P$, respectively.

According to the geometry of dumbbell-shaped tube, a self-locked energy-absorbing system can be assembled by stacking the dumbbell-shaped tubes in a staggered arrangement as shown in Fig. 1(c). This arrangement allows the column of dashed line dumbbell-shaped tubes to fit in the interspace between solid line dumbbell-shaped tubes in adjacent columns. Thus, when the system is impacted along the negative $y$-direction, each column of dumbbell-shaped tubes can intermesh with neighboring columns like a zip to suppress lateral splash in the $x$-direction. Therefore, this system can deform and absorb energy under impact loadings without additional constraints, such as welding, bolt or rivet fasteners. As a result, it breaks through the limitations of round-tube energy-absorbing system and greatly reduces labor and time costs during installation and removal.

## 3 Dumbbell-shaped tube system vs. round tube system

In this section, the numerical simulations are carried out to test the self-locking effect of the proposed energy-absorbing system constructed by dumbbell-shaped tubes. Comparison is made between the dynamic responses of the proposed dumbbell-shaped tube system and round tube system in three aspects: deformed configuration, impact force, and energy absorption capacity.

### 3.1 Finite element analysis

The numerical simulations are performed by the finite element method (FEM) using ABAQUS/Explicit. The main geometric parameters of the dumbbell-shaped tube and round tube systems are listed in Table 1. The dumbbell-shaped and round tubes are 100 mm in length and modeled using shell element S4R in ABAQUS with mechanical properties as follows: the density $\rho$=7830 kg/m$^3$, the elastic modulus $E$=207 GPa and the Poisson's ratio $v$=0.28. The isotropic hardening is taken into account by a constant tangent modulus of $E_{\tan}$=1.1 GPa, as

$$\sigma = \sigma_0 + E_{\tan}\left(\varepsilon - \frac{\bar{\sigma}}{E}\right) \quad (\sigma > \bar{\sigma}) \tag{1}$$

where $\sigma$, $\varepsilon$ and $\bar{\sigma}$ are equivalent stress, equivalent strain and equivalent yield stress, respectively. The Cowper-Symonds overstress power law is adopted to describe the strain rate dependence of the tube material, expressed as

$$\dot{\bar{\varepsilon}}^{pl} = C\left(\frac{\bar{\sigma}}{\sigma_0} - 1\right)^n \tag{2}$$

in which $\dot{\bar{\varepsilon}}^{pl}$ is the equivalent plastic strain rate, $\sigma_0$ the static yield stress, and $C$ and $n$ the material parameters. In our simulation, $\sigma_0$=322 MPa, $C$ = 40 and $n$ = 5 (Reid and Reddy, 1980).

The energy absorbing system is placed between a rigid striker and a fixed rigid supporting plate. The striker is either (1) a cylindrical striker, with a diameter of 126 mm and the axis parallel to the tube axes, to simulate the concentrated impact, or (2) a flat striker with a width of 410 mm, to simulate the uniformly-distributed impact. Both strikers are 118.33 kg and impact upon the tubes along the negative $y$-direction at an initial velocity of 4.43 m/s.

The "hard" contact interactions are set between all contacting surfaces, i.e. between the striker and tubes, the supporting plate and tubes, the adjacent tubes, and the self-contact of each tube. The friction coefficient in all simulations is 0 except that for round tube systems under uniformly-distributed impact loading, in which three levels of friction coefficient, $f = 0$, 0.05 and 0.1, are used to simulate the friction between round tubes. For the round tube system under concentrated impact loading, two rigid walls with a height of 75.6 mm are established to simulate the lateral boundary constraint for round tubes. All simulations are carried out in a gravitational field.

### 3.2 Concentrated impact

A cylindrical rigid striker is initially located at the top layer of dumbbell-shaped tubes, tangent to the upper surfaces of tube, as shown in Fig. 2(a). For comparison, round-tube energy absorbing systems with two different boundary conditions are employed: the one is free in the lateral direction as shown in Figs. 2(b), named as free round tube system, and the other is constrained in the lateral direction by two fixed rigid walls as shown in Figs. 2(c), named as constrained round tube system.

For the self-locked system, 33 dumbbell-shaped tubes are stacked in five layers by placing 7, 6, 7, 6 and 7 tubes from top to bottom, as shown in Fig. 2(a). The total height of the whole system is 71.4mm. Similarly, 70 round tubes make up an energy absorbing system by placing in four layers with 17, 18, 17 and 18 round tubes from top to bottom, with a total height 75.6mm, as shown in Figs. 2(b) and (c).

The configurations of above three cases at a normalized striker displacement of 0.7 are shown in Figs. 2(d), (e) and (f) respectively. Here the normalized displacement is defined by the ratio of total striker displacement $u$ to the total system height $H$, and 0.7 corresponds to a displacement of 50.0 mm for the self-locked system in Fig. 2(a) and 52.9 mm for the round tube systems in Figs. 2(b) and (c). Comparing the deformed configurations, it can be found that the dumbbell-shaped tubes can lock each other during impact and stay in their original positions, while the round tubes splash in lateral direction and most tubes fly far away from their original places. These results suggest that the proposed system can lock the tubes in their original places under impact loading.

Furthermore, the round tube system in Fig. 2(e) expands severely in the lateral direction, resulting in a very low force and thus small energy absorption. In Fig. 2(f), the scatter direction of the round tubes in the system with lateral constraint is unpredictable and thus the splashing is extremely dangerous and harmful.

To make a detailed investigation into the energy absorbing properties of the three different systems in Fig. 2, the force $F$ and the energy absorption $EA$ are studied and presented in Fig. 3 with respect to the normalized displacement of striker. The energy absorption $EA$ was obtained by integrating the impact force in Fig. 3(a) from 0 to 0.7 normalized displacement:

$$EA = \int_0^{0.7H} F \mathrm{d}u \qquad (3)$$

where $u$ is the striker displacement and $F$ is the simultaneous force.

From Fig. 3(a), the initial peak forces are observed in the curves of both round tube systems, and the system with lateral constraint has a higher initial peak force. Both curves are identical till 0.07 normalized displacement. After that, the force of free round tube system decreases sharply at 0.15 normalized displacement and then maintains in a very low level, which indicates that this system cannot effectively bear the impact force or absorb the impact energy. This concurs with the blue curve with solid triangles in Fig. 3(b), where the specific energy absorption is almost constant after an initial increase. However, the red curve with solid squares in Fig. 3(a) indicates that the force of constrained round tube system reaches its highest value at 0.16 normalized displacement and stays at a high level till 0.48 normalized displacement. Thus the energy absorption in Fig. 3(b) keeps increasing until the upper tubes splash out through the gaps between the fixed boundaries and the cylindrical striker.

Different from the round tube system, the force of self-locked system increases gradually and gently, without initial peak force, which means the self-locked system is better at protection in the initial impact stage. In addition, the energy absorption of self-locked system keeps increasing, and can reach and even exceed that of the constrained round tube system. Therefore, the self-locked system can mitigate the initial impact force and absorb impact energy effectively, without reduction in energy absorbing capability.

### 3.3 Distributed impact

A rigid loading plate with a width of 410 mm is established as the striker in ABAQUS/Explicit. It is initially located at the top of tubes, and just tangent to the upper surface of tubes as shown in Figs. 4(a) and (b). Frictionless contact conditions is set in the simulation of the self-locked system. For the round tube system, the friction coefficient $f$ between tube surfaces is set as three different levels, $f$ = 0, 0.05 and 0.1, which represent the effect of no friction, low friction and high friction between tubes, respectively.

The self-locked system consists of 4 layers of dumbbell-shaped tubes, with a total height of 58.8 mm and a total mass of 1.741 kg, as shown in Fig. 4(a). From the top to the bottom, each layer has 3, 4, 3 and 4 tubes, respectively. Similarly, 42 round tubes build up the round tube system by placing 10, 11, 10 and 11 tubes in four layers from the top to the bottom, with a total height of 75.6 mm and a total mass of 1.736 kg, as shown in Fig. 4(b).

Figs. 4 (d), (e), (f) show the deformed configurations of the above round tube systems at a normalized striker displacement of 0.7. It can be found that the tubes in round tube systems all spread under the impact no matter the friction between tubes is low or high. Thus boundary constraints and/or fasteners between tubes are definitely needed in the practical applications to prevent secondary damages. However, the dumbbell-shaped tubes intermesh with each other under impact and thus neighboring tubes can provide lateral constraints to each other, achieving the self-locking effect, as shown in Fig. 4(c). No lateral splashing is observed from the self-locked system.

Fig. 5(a) shows the force response versus the normalized displacement of striker. For round tube systems, an initial peak force appears at about 0.02 normalized displacement regardless of the magnitude of friction coefficient, then the force of round tube systems gradually decreases to zero due to the splashing of tubes. However, the friction between round tubes can slow the decrease because it retards the scattering of tubes. It can be observed from Fig. 5(a) that the round tube system with larger friction coefficient has higher force. At about 0.45 normalized displacement, the force of round tube systems all begins to rise because the tubes still under the flat striker starts to carry the impact loadings. In this stage, the friction coefficient plays an important role because it decides how many tubes can stay under the striker and how many tubes have been struck away. In contrast to round tube systems, the force of self-locked system increases gradually without significant peak forces, which is similarly to the response under concentrated impact discussed in section 3.2.

The energy response is shown in Fig. 5(b). Though all round tube systems have high energy absorption at early stages of impact due to a large initial peak of impact force, their energy absorption capacity is quickly exhausted after 0.1 normalized displacement as the round tubes are knocked away and the system is no longer able to carry impact loads. The proposed self-locked system, even if without friction between tubes, can still resist lateral splashing and gradually absorb the impact energy.

The splashing of round tubes is dangerous and reduces the energy absorption of the systems, especially for the systems with no friction or low friction, as shown in Fig. 4. Thus, the self-locked system is a safer alternative to the round tube systems.

## 4  Experimental validation of the self-locking effect

In this section, the impact experiment was carried out to validate the self-locking effect of the dumbbell-shaped tube system, and to verify the numerical model used in section 3 as well.

### 4.1 Experimental preparation

*4.1.1 Manufacturing of dumbbell-shaped tubes*

The dumbbell-shaped tube was manufactured by compressing a mild steel round tube with suitable dies and cores as shown in Fig. 6(a). The wall thickness and diameter of the round tube were 0.8 mm and 62.4 mm respectively. A manufactured dumbbell-shaped tube is shown in Fig. 6(b), and the average geometric parameters and mass of 14 dumbbell-shaped tube specimens are listed in Table 2.

*4.1.2 Impact experiment of the self-locked system*

The impact experiment was conducted using JSL-3000J drop hammer apparatus. The maximum impact energy is 3000J, the maximum drop height is 1.2 m and drop mass ranges from 70 to 300 kg. In this research, the drop height was 1.0 m and the drop mass was 118.33 kg.

The self-locked energy-absorbing system was 58.8 mm in height and composed of 14 dumbbell-shaped tubes as photographed in Fig. 7(a). In this picture, the dumbbell-shaped tubes were placed in four layers between two parallel thick plates and the impact was loaded on the upper plate by a striker. Each layer from top to bottom had 3, 4, 3 or 4 dumbbell-shaped tubes. Both plates were 7.63 kg, 20 mm in thickness, 410 mm in width and made of 1045 steel.

**4.2 Results and discussion**

*4.2.1 Validation of the self-locking effect*

Fig. 7(c) presents the experimental results of the deformed self-locked system at 0.37 normalized displacement. It can be found that the tubes in the top and the bottom layers deformed more severely than those in the middle layers. This phenomenon occurred because of the reflection and interference of stress wave. The compression wave was reflected as a compression wave when reaching an object with higher modulus. In this way, the compressive stress in the bottom and top layers was enhanced, while that in the middle layers was weakened.

The top layer deformed earlier than other layers under the impact loading. The flat plates of the dumbbell-shaped tubes in the top layer began to curve and touch the upper rigid plate. At the same time, the deformation zone gradually extended downward from the upper layer. Dumbbell-shaped tubes in the top, middle and bottom layers started to deform one layer after another while the top and bottom layers deform faster than the two middle layers. Due to the complex contact interactions in multi-tube crush, every single dumbbell-shaped tube deformed in coordination with adjacent tubes. Deformation of tubes at the free ends of the system was slightly smaller than those inside due to the lack of constraint. The impact energy was mainly absorbed by the plastic deformation of cylindrical shells rather than the parallel plates which almost remained flat after the impact.

Fig. 7(e) shows the deformed configuration after impact. It should be noted that all dumbbell-shaped tubes stay at their original positions even without the presence of lateral constraint. This proves the horizontal self-locking effect.

*4.2.2 Verification of the finite element model*

In Fig. 7(b), a finite element model of the self-locked system is developed according to the experiment. The mesh, constitutive relation, interaction and boundary conditions of the model are the same as those in section 3. The geometric parameters of the dumbbell-shaped tube are similar to the specimens and listed in Table 1. To simulate the experimental process, the upper plate with a mass of 7.63 kg is placed on the self-locked system and impacted by a rigid drop hammer striker with a mass of 118.33 kg and an initial velocity of 4.43 mm/s along the negative *y*-axis. Only translational movement in the *y*-direction is allowed for the striker.

The striker displacement $u$ verses time is plotted in Fig. 9. The predicted curve agrees with the experimental curve. The relative error in the maximum striker displacement is 7.5%. The configuration of self-locked system at 0.37 normalized striker displacement obtained from simulation is plotted in Fig. 7(d), which is consistent with that observed in experiment in Fig. 7(c). Therefore, the finite element model of the self-locked system under impact loading is valid.

## 5 Parameter study and optimal design of self-locked systems

The force and energy absorption capacity are important properties in the evaluation of an energy absorbing system. In order to improve the performance of the proposed self-locked system, the effect of geometric parameters of a single dumbbell-shaped tube is studied and suggestions to optimize the tube is given, and then an approach for the design of stacking arrangements of the system is suggested based on the investigation of the quasi-static compression of dumbbell-shaped tube arrays.

## 5.1 Parameter study of a single dumbbell-shaped tube

In order to achieve better performance, the effects of geometric parameters of the cross section of dumbbell-shaped tube are studied. The design parameters of the tube are demonstrated in Fig. 1(b), including the diameter $D$, the thickness $t$, the axis distance $A$ and the plate spacing $P$. Effects of generalized parameters $t/D$, $A/D$ and $P/D$ are analyzed to determine the relationship between the shape and energy absorption, and the size effect is investigated by examining the effect of variations in $D$ when $t/D$, $A/D$ and $P/D$ are constant.

To carry out the geometric parameter study, a finite element model is developed to study the mechanical responses of a dumbbell-shaped tube under static compressive loading. The simulation model is validated by quasi-static compression experiment.

### 5.1.1 Finite element model and experimental validation

FEM simulations are performed on the static compression of a single dumbbell-shaped tube by ABAQUS/Standard. The geometric parameters of dumbbell-shaped tube model are listed in Table 1. Because the axial tube length $L$ is much larger than the thickness $t$, the tube is assumed under plane strain condition as addressed in literature (DeRuntz and Hodge, 1963; Gupta et al., 2005; Reid and Reddy, 1978). The cross section of a dumbbell-shaped tube is modeled with plane strain element CPE4R, which includes 4-node bilinear reduced integration with hourglass control. The loading and supporting plates are idealized as rigid bodies since their structural stiffness is much larger than that of the thin-walled dumbbell-shaped tube.

In the simulation, a single tube is placed between two parallel rigid plates, as shown in Fig. 9(a). The displacement load along the negative $y$-direction is applied to the upper rigid plate and all degrees of freedom of the supporting rigid plate is constrained. Due to the symmetry, only a half of the dumbbell-shaped tube is modeled. Symmetric boundary condition is applied to the symmetric planes to eliminate antisymmetric displacement and rotation. The contact properties between surfaces are set as "surface to surface", "frictionless" and "hard" contact.

To study the convergence of the simulation model, the force-displacement curves of various mesh densities are plotted in Fig. 10. It can be seen that all curves are close to each other and the force of the model with element size 0.2 mm agrees well with the models with smaller element sizes. Therefore, the element size 0.2 mm is fine enough for this study, which corresponds to a mesh density of 4 elements through thickness. A mesh density with at least 4 elements through thickness is thus used in the subsequent FEM simulations.

The force-displacement curve of the dumbbell-shaped tube obtained from simulation is depicted in Fig. 11. The normalized displacement in the figure is defined as the ratio of the upper plate displacement $u$ to the cylinder shell diameter $D$. The simulation curve in Fig. 11 can be divided into three stages: elastic stage, hardening stage and flattening stage. In the first stage, the force increases proportionally with the plate displacement since the compression is dominated by the elastic deformation. The dumbbell-shaped tube absorbs only a little energy during this short elastic process. As the compression progresses, the two flat plates of tube get closer and closer. When they are in complete contact with each other, as shown in Fig. 9(b), the curve reaches the first inflection point and then goes into the second stage. In this stage, the middle of plates begins to separate as shown in Fig. 9(c). The force continues to increase with a more gentle slope and the impact energy is efficiently absorbed by the plastic deformation. The third stage begins at the second inflection point where both plates touch the rigid flat plates as shown in Fig. 9(d). In this stage, the dumbbell-shaped tube is gradually flattened by the rigid plates as shown in Fig. 9(e) and the force curve becomes steeper, which means even a small displacement increment can lead to a large increase in force. The tube is severely damaged and no longer suitable for energy absorption or impact protection.

It can be found that the energy is mainly absorbed in the second stage by investigating the area under the force curve in Fig. 15. Although the energy absorption in the third stage is quite large, it is dangerous to design the energy absorber with the energy absorption of the entire third stage taken into account, because the sharply increasing force suggests the exhaustion of energy absorption capacity and the tube might be totally flattened at any time in this stage. Therefore, the energy absorption capacity is calculated at 0.7 normalized displacement as in literature (Baroutaji et al., 2015). The specific energy absorption *SEA* is defined as the energy absorption per mass $m$

$$SEA = \frac{EA}{m} \tag{4}$$

where the energy absorption *EA* is obtained by integrating the simultaneous force *F* from the displacement $u=0$ to $u=0.7D$.

$$EA = \int_0^{0.7D} F \mathrm{d}u \tag{5}$$

The specific energy absorption *SEA* of a dumbbell-shaped tube is 677.8 J/kg, as listed in Table 3.

To validate the simulation model, three dumbbell-shaped tube specimens were tested by the CRIMS-100KN test machine with a loading speed of quasi-static compression of 1.0 mm/min. The dumbbell-shaped tube was placed between two thick steel plates as photographed in Fig. 9(f). The geometric parameters of dumbbell-shaped tube specimens are listed in Table 2.

Typical deformation modes predicted by the FEM simulation are compared with experiment as shown in Fig. 9. It can be seen that the deformed shapes of dumbbell-shaped tube in Figs. 9(a)-(e) all agree well with experimental results in Figs. 9(f)-(j).

The force response of the upper loading plate is plotted in Fig. 11 versus the normalized displacement. The force predicted by the FEM simulation agrees well with the experimental result. The specific energy absorption of experimental and simulation results are listed in Table 3. The relative error in specific energy absorption is only 3.1%.

From above, the predicted force and energy responses of FEM simulation are consistent with experiment. Therefore, the effects of the geometric parameters on the properties of the dumbbell-shaped tube are analyzed by FEM simulations for efficiency.

*5.1.2 Effect of the geometry on the force response*

In the crashworthiness design, the criteria to evaluate the performance of an energy absorbing system are commonly established by measuring the force *F*, the specific energy absorption *SEA* and the crushing load efficiency *CLE*. Therefore, *F*, *SEA* and *CLE* are chosen in this paper to describe the performance of a single dumbbell-shaped tube.

Fig. 12 shows the compressive force-displacement curves of dumbbell-shaped tubes with various geometric parameters: the thickness to diameter ratio *t/D*, the ratio of axis distance to diameter *A/D*, the ratio of plate spacing to diameter *P/D* and the diameter *D*. It can be observed that the force curves in Figs. 12 (a), (b), (c) and (d) all present three-stage increasing with two inflection points as discussed in section 5.1.1. For all the cases, in the first stage, the dominant deformation is elastic deformation and thus the force rises linearly with deformation until the first inflection point, at which the two flat plates are in complete contact. The dumbbell-shaped tube starts to yield at the second stage and the slope of force-deformation curve is reduced. After the second inflection point, the flat plates touches the rigid loading plates, and the dumbbell-shaped tube is soon flattened as the curve reaches the third stage and increases rapidly. The similar shape of curves implies that the typical deformation modes of dumbbell-shaped tubes are the same although the cross section geometries are different.

In Fig. 12(a), the force increases with the thickness to diameter ratio $t/D$. It is obvious that the thicker tubes can carry larger loads throughout the compression. This effect agrees with the effect of wall thickness on crush load, as addressed by DeRuntz and Hodge (1963) and Reid and Reddy (1978).

In Fig. 12(b), the responses of dumbbell-shaped tubes with different axis distance to diameter ratios $A/D$ are identical in most parts except for that after the first inflection and at the second inflection. At the first inflection, the flat plates connecting the cylindrical shells have been completely attached to each other. The interactions between these plates are normal contact forces concentrated at both ends of these plates. As the compression progresses, the plates start to bend convexly to resist the increasing of compressive loads. Since the dominant deformation mode of these plates are bending, shorter plates have a larger curvature at the same deflection, which generates larger bending moment, imposes stricter constraint on the deformation of cylindrical shells and eventually results in higher structural stiffness and higher reaction force at the beginning of the hardening stage. The difference in $A/D$ also affects the position of the second inflection. Deflections in shorter plates of dumbbell-shaped tube are smaller, which consequently postpone the contact with the rigid plates.

Fig. 12(c) presents the effect of plate spacing to diameter ratio $P/D$ on the force response of tubes. It is clear that there are two spots where the curves divert from each other. The first spot indicates the first contact between two flat plates. The curve with $P/D = 0.1$ separates from other curves in the beginning, because it represent the response of a dumbbell-shaped tube whose flat plates are initially in contact with each other. The curves with smaller plate spacing increases faster and goes into the hardening stage earlier than those with larger spacing, because their plates are easier to be compressed to contact each other completely. After the first separation, the force curves gradually converge as a result of similar deformation configurations until the second inflection point. The plates with smaller spacing can sooner be bent and touch the rigid loading plates.

The effect of diameter $D$ is displayed in Fig. 12(d). In Fig. 12(d), the dumbbell-shaped tubes with larger diameters have larger forces. This effect might look inconsistent with some previous literature (Baroutaji et al., 2015; Gupta et al., 2005; Reid and Reddy, 1978) where the crush force decreases with increasing of round tube diameter. It should be noted that in their studies the thickness is an independent variable of diameter, whereas in this work the thickness to diameter ratio $t/D$ is a constant, which means the thickness $t$ depends on the diameter. Although larger tubes have a longer moment arm that could reduce the load, its effect is counteracted by the increase in load carrying capacity caused by changes in thickness.

In summary, the magnitude of force is mainly influenced by the thickness to diameter ratio $t/D$ and the diameter $D$. Therefore, in applications where the impact force needs to be reduced, dumbbell-shaped tubes with smaller thickness to diameter ratio $t/D$ and smaller diameter $D$ should be employed. The geometry of dumbbell-shaped tube affects not only the magnitude of force, but also the shapes of force-displacement curves. Different axis distance to diameter ratio $A/D$ and plate spacing to diameter ratio $P/D$ can be employed to control the position of the first and second inflection points and obtain a desirable force response according to practical needs.

*5.1.3 Effect of the geometry on the specific energy absorption*

The effects of the generalized parameters, $t/D$, $A/D$ and $P/D$, and the diameter $D$ on the specific energy absorption *SEA* are presented in Fig. 13.

As observed from Fig. 13(a), the specific energy absorption is an increasing function of thickness to diameter ratio $t/D$, which suggests that the thicker tubes are more efficient than the thinner ones in energy absorption. This is consistent with previous studies on round tube (Choi et al., 1986; Gupta et al., 2005; Reid and Reddy, 1978).

In Figs. 13(b) and (c), a slight decreasing trend in specific energy absorption can be noted with the increase in the axis distance to diameter ratio from *A/D*=2 to *A/D*=5 and the plate spacing to diameter ratio from *P/D*=0.1 to *P/D*=0.4. Compared to *t/D*, the effects of *A/D* and *P/D* are less significant.

It is observed in Figs. 13(a), (b) and (c) that the curves of the dumbbell-shaped tubes with different diameters are very close to each other, which means that the specific energy absorption is dominated by the shape but not significantly affected by the tube size. Besides, Fig. 13(d) reveals that the increase in tube size from *D*=30mm to *D*=70mm only leads to a small variation of 5% in specific energy absorption.

In order to design a dumbbell-shaped tube with larger energy absorption capacity, the thickness to diameter ratio *t/D* should be as large as possible while the plate spacing to diameter ratio *P/D* and axis distance to diameter ratio *A/D* should be minimized. The effect of diameter can be ignored. Since the energy absorption capacity is not the only concern in anti-impact design, these parameters cannot be maximized or minimized infinitely. For example, increasing *t/D* leads to an increase in the impact force as presented in Fig. 16(a), and the geometry of the dumbbell-shaped tube requires that the plate spacing *P* is more than twice of the wall thickness *t* since the thickness of two flat plates is included in the plate spacing as shown in Fig. 1(b).

*5.1.4 Effect of the geometry on the crushing load efficiency*

Crushing load efficiency *CLE* describes the consistency and uniformity of collapse load. Due to the monotonically increasing nature of the force-displacement curve shown in Fig. 12, the maximum force $F_{max}$ is picked at *u*=0.7*D*. The crushing load efficiency *CLE* is then defined as (Huang and Wang, 2009)

$$CLE = \frac{F_{mean}}{F_{max}} \qquad (6)$$

where $F_{mean}$ is the mean crushing force, expressed as

$$F_{mean} = \frac{1}{0.7D}\int_0^{0.7D} F \mathrm{d}u \qquad (7)$$

It can be seen from Fig. 14(a) that the crushing load efficiency *CLE* rapidly decreases with the increase in thickness to diameter ratio *t/D*. This means that the thinner tubes are preferable to thicker ones for carrying load with higher efficiency.

The crushing load efficiency *CLE* versus the axis distance to diameter ratio *A/D* is shown in Fig. 14(b). The *CLE* increases with *A/D* and the slope decreases with *A/D*. As can be seen from Fig. 12(b), it is easy to spot the second inflection point for the force curve with *A/D* less than 3 because it has an obvious increase in the slope of force curve at the second inflection point which might be responsible for the low *CLE*. When *A/D* is more than 3, the second inflection becomes less noticeable and the transition of load between the second and third stages becomes smooth. This suggests that the optimal value of *A/D* for a maximized crushing load efficiency could be around or more than 3.

The curve of *CLE* versus the plate spacing to diameter *P/D* shows a negative slope in Fig. 14(c), which indicates that the dumbbell-shaped tubes with smaller plate spacing have smoother force responses. This is validated by the force curve in Fig. 12(c), in which the curves with a smaller *P/D* have a longer hardening stage and thus have a higher *CLE*.

The tube size does not have much influence on the crushing load efficiency as shown in Fig. 14(d). The curves is nearly horizontal, and the fluctuation in *CLE* is trivial when compared to Figs. 14(a), (b) and (c). It is reasonable to ignore the effect of tube size.

Therefore, the crushing load efficiency *CLE* is dominated by the shape, especially by the thickness to diameter ratio *t/D*, but not significantly affected by the tube size. Adjustment could be made on *A/D* and *P/D* in addition to *t/D* to achieve the desirable crushing load efficiency.

*5.1.5 Suggestions on the design of single dumbbell-shaped tubes*

In the design of an energy-absorbing element, it is important to take both specific energy absorption and force responses into account. Although the optimal values of geometric parameters with respect to different criteria can be quite contradictory, the following suggestions on the design of dumbbell-shaped tube are provided according to application preference:

1. The most effective way to increase specific energy absorption is to increase the

thickness to diameter ratio *t*/*D* but meanwhile the force is increased as well.

2. The force during compression can be reduced by choosing dumbbell-shaped tubes with smaller diameter to thickness ratio *t*/*D* or smaller diameter *D*. The axis distance to diameter ratio *A*/*D* and the plate spacing to diameter ratio *P*/*D* can affect the shape of the force-deformation curve but can hardly change the magnitude of force.

3. To make the distribution of force during compression more uniform, i.e. to increase crushing load efficiency, it is suggested to lower the thickness to diameter ratio *t*/*D* and increase the axis distance to diameter ratio *A*/*D* if it less than 3. Other two parameters, *P*/*D* and *D* have less impact on the crushing load efficiency.

**5.2 Optimal design of self-locked systems constructed by multi-row tubes**

From section 5.1, the optimal geometry of a single dumbbell-shaped tube can be determined via energy and force requirements. However, the performance of a self-locked system is affected not only by the properties of each individual dumbbell-shaped tube but more importantly by the stacking arrangements of these tubes. On one hand, dumbbell-shaped tubes should be assembled in a staggered fashion as in Fig.1(c), which can be considered as a precondition to achieve lateral constraint between tubes. On the other hand, what still needs to be studied is the stacking profile of dumbbell-shaped tubes. In this work, rectangularly-stacked self-locked systems compressed between two rigid flat plates are studied due to simplicity.

With the rectangular stacking arrangement, the stacking configuration can be described with two parameters: (1) the number of layers $n_{\text{layer}}$, and (2) the number of tubes in width $n_{\text{width}}$, which is defined by the maximum number of tubes in a layer. For example, the model shown in Fig. 15(a) represents a system with $n_{\text{layer}}=10$ and $n_{\text{width}}=7$. The effects of these two stacking parameters on the energy absorption and force responses of the system under quasi-static compression are studied by ABAQUS/Explicit using S4R shell elements. The simulation model in section 3 is used here.

*5.2.1 Effect of number of layers*

To study the effect of layer number $n_{layer}$ on the force and energy absorption of the self-locked system, the number of tubes in width is fixed to $n_{width} = 7$ and the number of layers $n_{layer}$ ranges from 3 to 31.

The force response curves versus normalized displacment are shown in Fig. 16(a). The shapes of the curves for different number of layers are similar. All the force curves start with a small slope, and then rise sharply after 0.1 normalized displacement. At about 0.17 normalized displacement, they start to separate from each other and the systems with more layers have higher force responses, and the free cylindrical ends of the dumbbell-shaped tubes in neighboring layers contact with each other at this time, as shown in Fig. 15(b). When the dumbbell-shaped tubes at the bottom and top layers are flattened as in Fig. 15(c), the force curve reaches an inflection point. A plateau phase is then observed. The slope of force in the plateau phase is smaller for systems with more layers. A similar plateau phase has been reported by Shim and Stronge (1986) in the quasi-static compression of rectangularly-stacked aluminum and brass round tubes. Following the plateau is a final increase which indicates the exhaustion of energy absorption capacity as plotted in Fig. 15(d).

The energy absorption at 0.7 normalized displacement is shown in Fig. 16(b). It can be found that the energy absorption increases linearly with the number of layers.

*5.2.2 Effect of tube number in width*

To study the effect of the number of tubes in width on the force and energy absorption of the self-locked system, the number of layers is fixed to $n_{layer} = 10$ and the number of tubes in width $n_{width}$ varies from 2 to 12.

The quasi-static responses of self-locked system with different widths are shown in Figs. 17(a) and (b). Fig. 17(a) presents the force-displacement curves of self-locked system with different widths. The increases in width lead to increases in the magnitude of compressive forces, and the shapes of these curve are similar. In Fig. 17(b), a linear relation is found between the energy absorption and the number of tube in width, same as that in Fig. 16(b).

*5.2.3 A guideline on stacking arrangement design*

It can be observed in Fig. 16 that the force curves are close to each other for the self-locked systems with different layers. The systems with more layers have higher forces and the variation in maximum forces is about 17.5% when the number of layers increases from 3 to 31. Therefore, the influence of the number of layers on the force responses is limited. However, the self-locked systems with more layers have of a longer stroke of compression. This implies that due to the prolonged displacement, the increase in force is milder and the energy absorption capacity is larger. In conclusion, the rectangularly-stacked self-locked system with more layers is more favorable in impact attenuation and protection of occupants.

The influence of the width of rectangularly-stacked self-locked system on the compressive load is different from that of the number of layers. An overview of load history is displayed in Fig. 17(a). The self-locked systems with more tubes in width have a higher force during compression, because the load carrying capacity is increased by additional columns of dumbbell-shaped tubes.

Energy absorption responses of self-locked systems with various layer number and width are summarized in Fig. 18. In spite of the different stacking arrangements, the energy absorption curves with respect to the total number of tubes coincides on a same straight line, which indicates that the energy absorption can be expressed by a linear function of the total number of tubes $n_{total}$. The linear expression is proposed here as

$$EA = EA_{tube} \times n_{total} \tag{8}$$

where $EA$ is the energy absorption of the system, and $EA_{tube} = 83.2J$ is the energy absorption of a dumbbell-shaped tube obtained in Section 5.1. Fig. 18 shows the proposed linear expression agrees well with the simulation results.

From above analysis, the energy absorption increases linearly with both the number of layers and the number of tubes in width, while the force is mainly affected by the number of tubes in width. Accordingly, a general guideline on the design of a rectangularly-stacked self-locked system is provided, as illustrated in the flow chart Fig. 19.

Using Eq. (8), it is easy to determine the minimum total number of dumbbell-shaped tubes according to the impact energy. Meanwhile, the number of tubes in width can be determined by the requirement of the maximum force as well as other factors like the size of the striker concerning the safety of occupants. With the minimum total number of dumbbell-shaped tubes and the number of tubes in width, the layer number of the rectangularly-stacked self-locked system is determined.

# 6 Conclusion

A novel self-locked energy absorbing system is proposed and studied by both experiments and simulations in this paper. The following conclusions on the self-locked system are drawn.

1. The proposed self-locked system is able to prevent lateral splash from impact loadings even without any boundary constraints or fasteners between dumbbell-shaped tubes. This self-locking effect is validated by both impact experiment and FEM simulations.

2. Compared to the round-tube system with lateral constraints, the proposed self-locked system can lead to a more gentle and more gradual increase in the reaction force, without loss of energy-absorbing capacity.

3. The geometric parameters of a single dumbbell-shaped tube and the stacking arrangement of multiple-tube systems are studied and optimized. A guideline on designing a self-locked energy-absorbing system is provided for practical applications.

Based on this work, future research could be carried out on designing a system with self-locking effect in more directions. Multi-directional self-locking systems can be obtained by a three dimensional design on the geometry of the tubes.

## Acknowledgments

Supports by the National Natural Science Foundation of China (Nos. 11202012 and 11472027) and the Program for New Century Excellent Talents in University (No. NCET-13-0021) are gratefully acknowledged.

response of composite tubes. Compos. Struct. 91, 222–228.

Johnson, W., Mamalis, A.G., 1978. Crashworthiness of vehicles. Mechanical Engineering Publications Limited, London.

Jones, N., 2011. Structural impact. Cambridge university press.

Jones, N., Okawa, D.M., 1976. Dynamic plastic buckling of rings and cylindrical shells. Nucl. Eng. Des. 37, 125-147.

Jones, N., Wierzbicki, T., 2010. Structural Crashworthiness and Failure: Proceedings of the Third International Symposium on Structural Crashworthiness held at the University of Liverpool, England, 14-16 April 1993. CRC Press.

Knoell, A.C., Wilson, A.H., 1978. Vehicular impact absorption system. US Patent 4,118,014.

Langseth, M., Hopperstad, O.S., 1996. Static and dynamic axial crushing of square thin-walled aluminium extrusions. Int. J. Impact Eng. 18, 949-968.

Lindberg, H., Kennedy, T., 1975. Dynamic plastic pulse buckling beyond strain-rate reversal. J. Appl. Mech. 42, 411-416.

Lindberg, H.E., 1965. Dynamic plastic buckling of a thin cylindrical shell containing an elastic core. J. Appl. Mech. 32, 803-812.

Lindberg, H.E., Florence, A.L., 1987. Dynamic pulse buckling: theory and experiment. Springer Science & Business Media.

Lu, G., Yu, T., 2003. Energy absorption of structures and materials. Woodhead Publishing Ltd, Cambridge.

Mills, N.J., Wilkes, S., Derler, S., Flisch, A., 2009. FEA of oblique impact tests on a motorcycle helmet. Int. J. Impact Eng. 36, 913-925.

Morris, E., Olabi, A.-G., Hashmi, M., 2006. Analysis of nested tube type energy absorbers with different indenters and exterior constraints. Thin-Walled Struct. 44, 872-885.

Morris, E., Olabi, A., Hashmi, M., 2007. Lateral crushing of circular and non-circular tube systems under quasi-static conditions. J. Mater. Process. Technol. 191, 132-135.

Neves, R., Micheli, G., Alves, M., 2010. An experimental and numerical investigation on tyre impact. Int. J. Impact Eng. 37, 685-693.

Niknejad, A., Elahi, S.A., Liaghat, G.H., 2012. Experimental investigation on the lateral compression in the foam-filled circular tubes. Mater. Des. 36, 24–34.
27

**Table caption list**

**Table 1** The parameters of dumbbell-shaped tube and round tube systems for different impact loading cases

**Table 2** The geometry of dumbbell-shaped tube specimens

**Table 3** The specific energy absorption of a single dumbbell-shaped tube (J/kg)

**Figure caption list**

**Figure 1**  Schematic of (a) a dumbbell-shaped tube, (b) the cross section of a dumbbell-shaped tube and (c) the self-locked energy-absorbing system.

**Figure 2**  Energy-absorbing systems under concentrated impact loading located upon a single tube: (a) self-locked system, (b) round tube system without lateral boundary constraint and (c) round-tube system with lateral boundary constraint before impact, as well as their configurations (d)-(f) at 0.7 normalized displacement.

**Figure 3**  Comparison between the self-locked system and the round tube systems with impact located upon a single tube: (a) force and (b) energy absorption.

**Figure 4**  Energy-absorbing systems under uniformly-distributed impact loading: (a) self-locked system and (b) round tube system before impact, as well as their configurations (c)-(f) at 0.7 normalized displacement with friction coefficients of 0, 0.05 and 0.1.

**Figure 5**  Comparison between the self-locked system and the round tube systems with uniform distributed impact: (a) force and (b) energy absorption.

**Figure 6**  Manufacture of a dumbbell-shaped tube: (a) manufacturing of a dumbbell-shaped tube and (b) the specimen.

**Figure 7**  Impact experiment and simulation: (a) experimental set-up, (b) the finite element model, (c)-(d) the self-locked system at 0.37 normalized displacement in experiment and simulation and (e) the self-locked system after impact experiment.

**Figure 8**  Striker displacement versus time of self-locked system under impact loading.

**Figure 9**  Compression of a dumbbell-shaped tube between two flat plates: (a)-(e) static simulation and (f)-(j) quasi-static experiment.

**Figure 10**  Convergence study of the finite element model.

**Figure 11**  Force response of a dumbbell-shaped tube under compression, in which the deformed configurations obtained from simulation and experiment at points (a)-(e) correspond to Figs. 9(a)-(e) and 9(f)-(j), respectively.

**Figure 12**  Force responses of dumbbell-shaped tubes with different (a) thickness to diameter ratio t/D, (b) axis distance to diameter ratio A/D, (c) plate spacing to diameter ratio P/D and (d) diameter D.

**Figure 13**  Specific energy absorption of dumbbell-shaped tubes with different (a) thickness to diameter ratio t/D, (b) axis distance to diameter ratio A/D, (c) plate spacing to diameter ratio P/D and (d) diameter D.

**Figure 14**  Crushing load efficiency of dumbbell-shaped tubes with different (a) thickness to diameter ratio t/D, (b) axis distance to diameter ratio A/D, (c) plate spacing to diameter ratio P/D and (d) diameter D.

**Figure 15**  Compressive simulation of a symmetric self-locked system with $n_{\text{layer}}=10$ and $n_{\text{width}}=7$ at normalized displacements of (a) u/H=0, (b) u/H=0.17, (c) u/H=0.25 and (d) u/H=0.7.

**Figure 16**  Effect of number of layers on (a) force and (b) energy absorption.

**Figure 17**  Effect of number of tubes in width on (a) force and (b) energy absorption.

**Figure 18**  Energy absorption versus the total number of tubes.

**Figure 19**  A general guideline on the design of a rectangularly-stacked self-locked system.

# Tables

**Table 1**  The parameters of dumbbell-shaped tube and round tube systems for different impact loading cases

| Tube type and parameters (mm) | Loading | Tube numbers in each layer | System parameters | | |
|---|---|---|---|---|---|
| | | | $H$ (mm) | $W$ (mm) | $M$ (kg) |
| Dumbbell-shaped tube 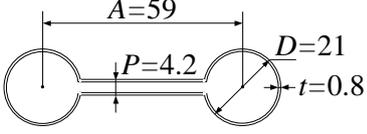 $A=59$, $P=4.2$, $D=21$, $t=0.8$ | Concentrated | 7/6/7/6/7 | 71.4 | 560 | 4.105 |
| | Distributed | 3/4/3/4 | 58.8 | 320 | 1.741 |
| Round tube 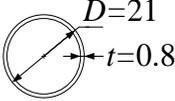 $D=21$, $t=0.8$ | Concentrated | 17/18/17/18 | 75.6 | 378 | 2.893 |
| | Distributed | 10/11/10/11 | 75.6 | 231 | 1.736 |

**Table 2** The geometry of dumbbell-shaped tube specimens

|  | Specimen | $D$ (mm) | $A$ (mm) | $t$ (mm) | $P$ (mm) | $L$ (mm) |
|---|---|---|---|---|---|---|
| Impact experiment | I-1 | 21.26 | 59.87 | 0.78 | 4.77 | 100.52 |
|  | I-2 | 21.29 | 59.98 | 0.78 | 5.25 | 100.62 |
|  | I-3 | 21.30 | 59.83 | 0.78 | 4.92 | 100.21 |
|  | I-4 | 21.29 | 60.03 | 0.78 | 5.22 | 100.48 |
|  | I-5 | 21.06 | 60.21 | 0.78 | 4.87 | 100.33 |
|  | I-6 | 21.14 | 60.03 | 0.78 | 4.61 | 99.92 |
|  | I-7 | 21.16 | 59.99 | 0.78 | 4.99 | 100.45 |
|  | I-8 | 21.03 | 60.38 | 0.78 | 5.13 | 99.84 |
|  | I-9 | 21.00 | 60.14 | 0.78 | 4.57 | 99.68 |
|  | I-10 | 20.97 | 60.41 | 0.78 | 4.71 | 98.98 |
|  | I-11 | 21.18 | 60.14 | 0.78 | 5.05 | 100.63 |
|  | I-12 | 21.28 | 59.86 | 0.78 | 4.91 | 99.30 |
|  | I-13 | 21.11 | 60.18 | 0.78 | 5.04 | 99.78 |
|  | I-14 | 21.17 | 60.04 | 0.78 | 5.08 | 100.56 |
|  | Average | 21.16 | 60.07 | 0.78 | 4.93 | 100.09 |
| Quasi-static experiment | Q-1 | 21.17 | 59.93 | 0.78 | 5.25 | 100.00 |
|  | Q-2 | 21.02 | 60.12 | 0.77 | 4.81 | 100.10 |
|  | Q-3 | 21.07 | 59.54 | 0.78 | 4.40 | 98.30 |

**Table 3** The specific energy absorption of a single dumbbell-shaped tube (J/kg)

| Quasi-static experiment | | | | FEM Simulation |
|---|---|---|---|---|
| No.1 | No.2 | No.3 | Average | |
| 674.9 | 704.4 | 719.4 | 699.6 | 677.8 |

# Figures

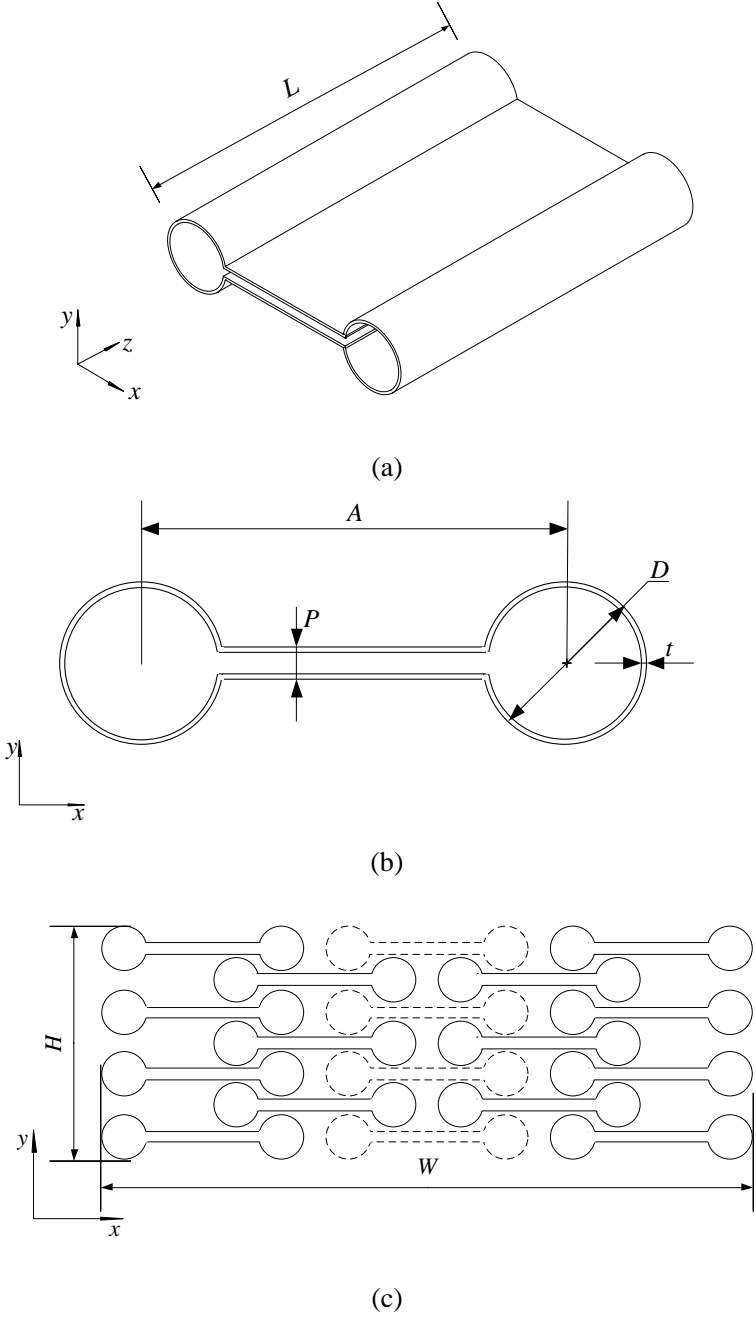

(a)

(b)

(c)

**Figure 1** Schematic of (a) a dumbbell-shaped tube, (b) the cross section of a dumbbell-shaped tube and (c) the self-locked energy-absorbing system.

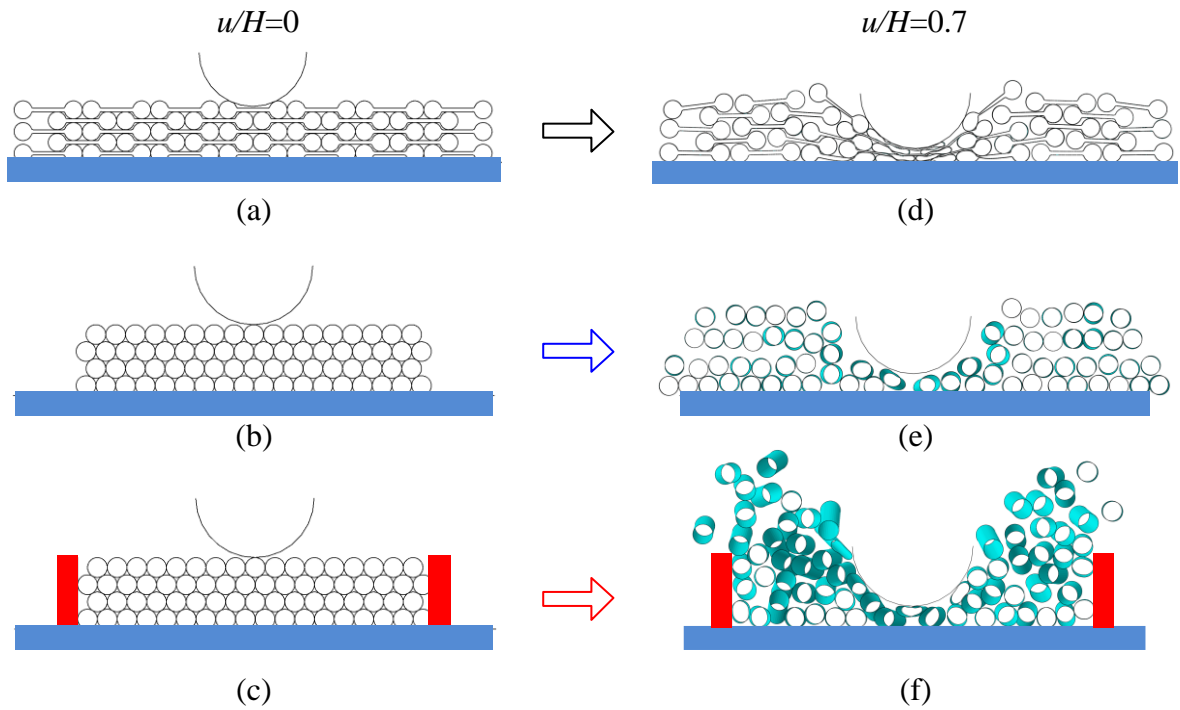

**Figure 2**  Energy-absorbing systems under concentrated impact loading located upon a single tube: (a) self-locked system, (b) round tube system without lateral boundary constraint and (c) round-tube system with lateral boundary constraint before impact, as well as their configurations (d)-(f) at 0.7 normalized displacement.

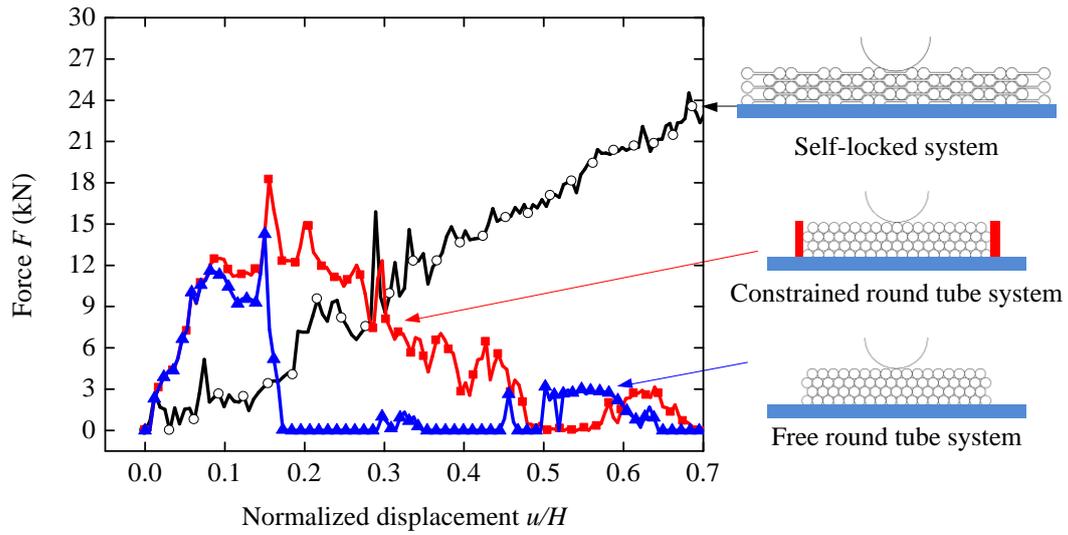

(a)

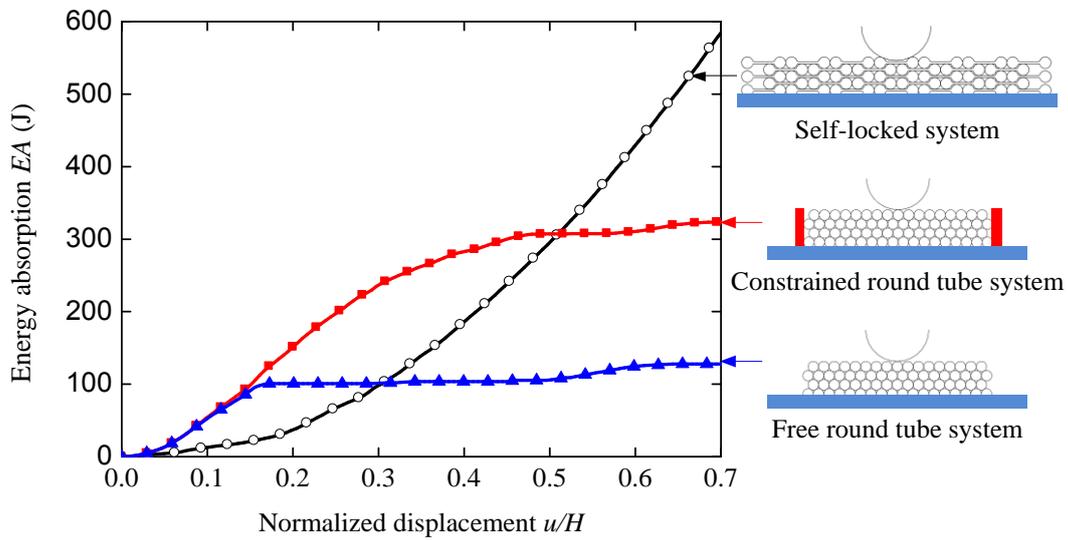

(b)

**Figure 3**  Comparison between the self-locked system and the round tube systems with impact located upon a single tube: (a) force and (b) energy absorption.

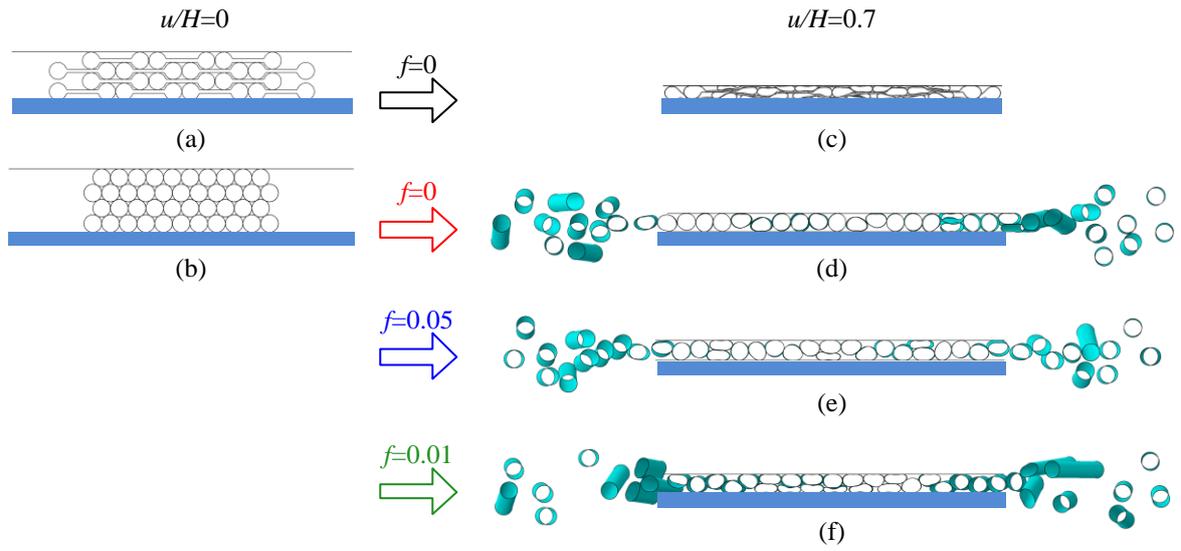

**Figure 4** Energy-absorbing systems under uniformly-distributed impact loading: (a) self-locked system and (b) round tube system before impact, as well as their configurations (c)-(f) at 0.7 normalized displacement with friction coefficients of 0, 0.05 and 0.1.

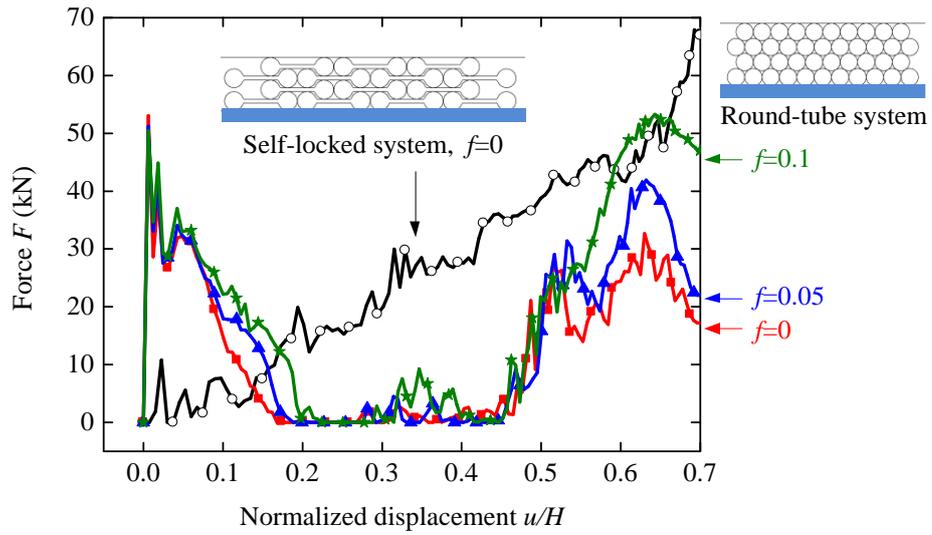

(a)

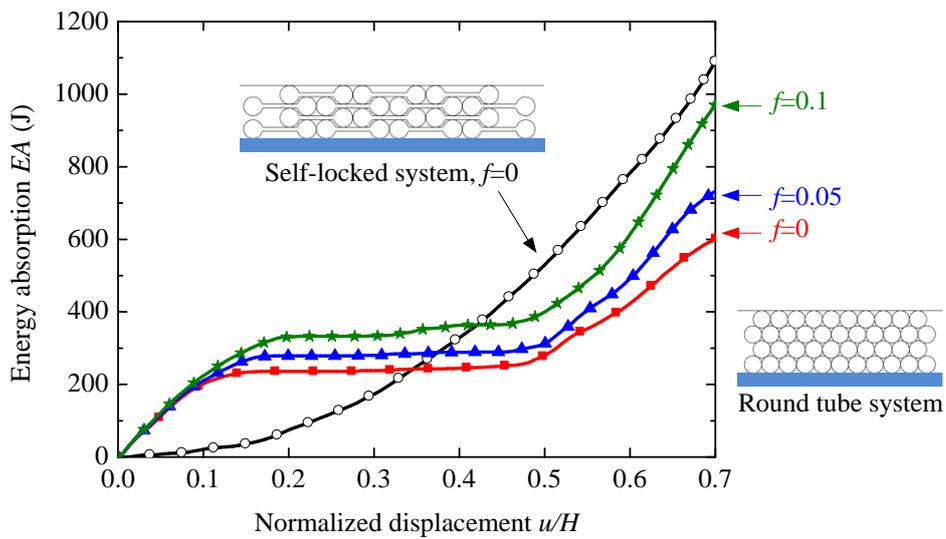

(b)

**Figure 5**  Comparison between the self-locked system and the round tube systems with uniform distributed impact: (a) force and (b) energy absorption.

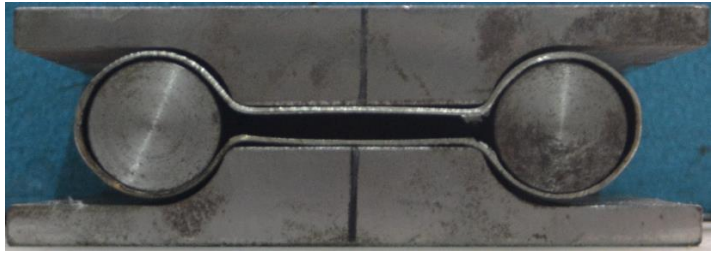

(a)

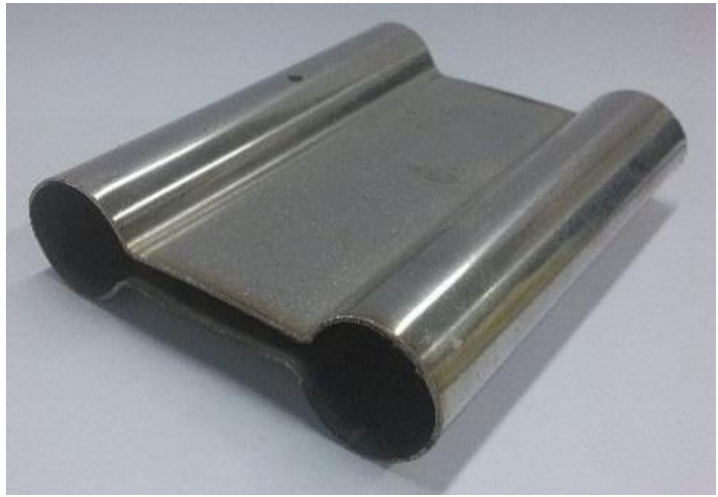

(b)

**Figure 6** Manufacture of a dumbbell-shaped tube: (a) manufacturing of a dumbbell-shaped tube and (b) the specimen.

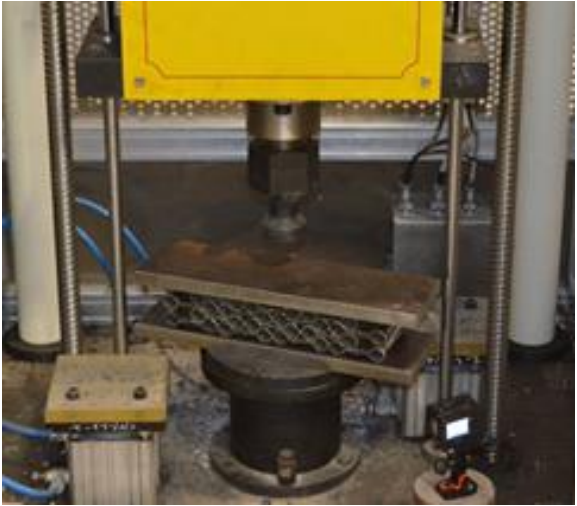 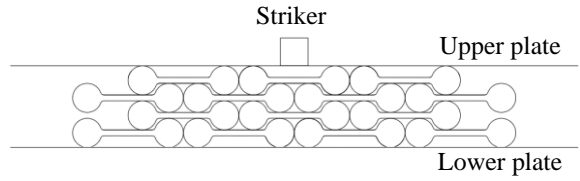

(a) (b)

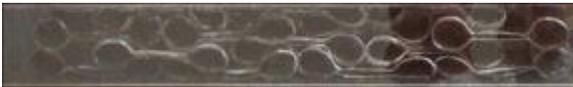 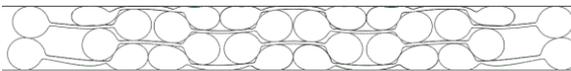

(c) (d)

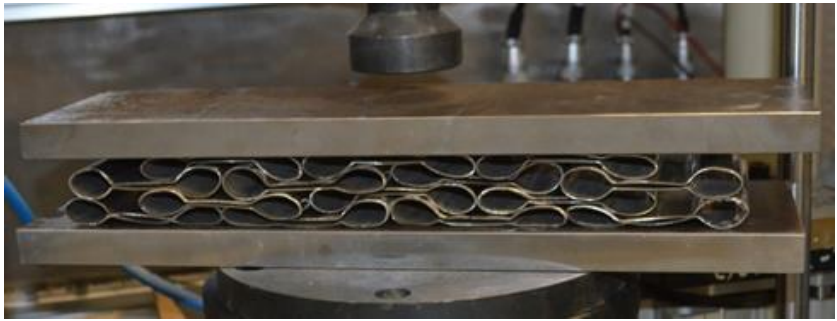

(e)

**Figure 7** Impact experiment and simulation: (a) experimental set-up, (b) the finite element model, (c)-(d) the self-locked system at 0.37 normalized displacement in experiment and simulation and (e) the self-locked system after impact experiment.

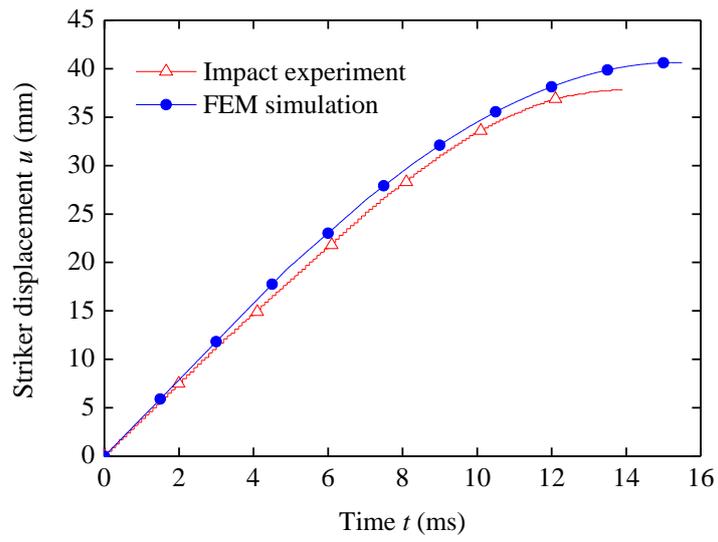

**Figure 8**   Striker displacement versus time of self-locked system under impact loading.

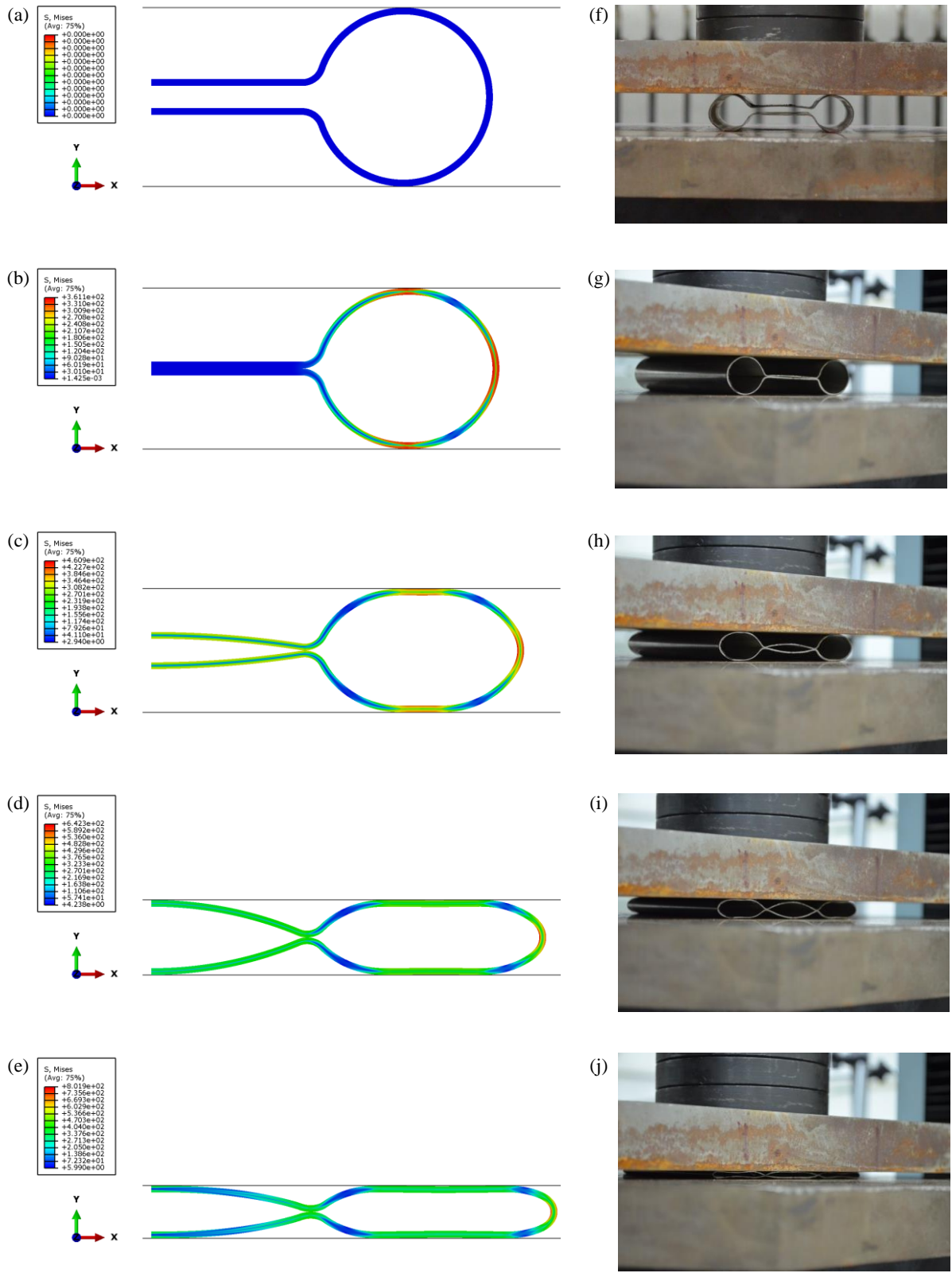

**Figure 9** Compression of a dumbbell-shaped tube between two flat plates: (a)-(e) static simulation and (f)-(j) quasi-static experiment.

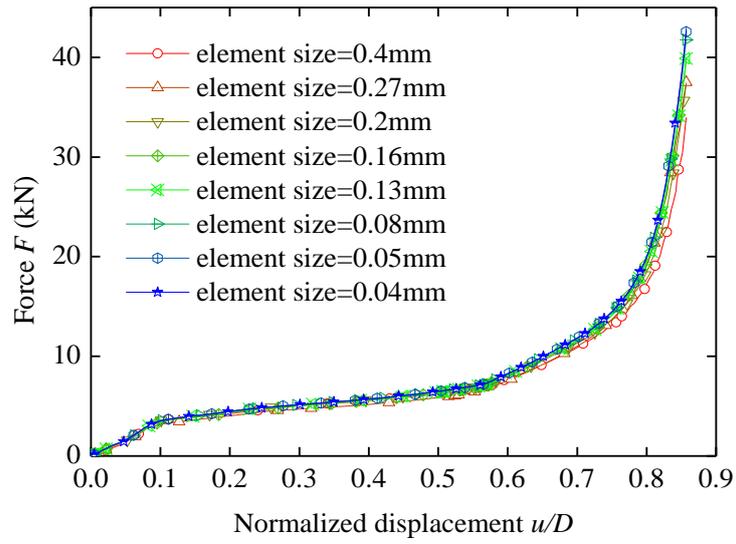

**Figure 10** Convergence study of the finite element model.

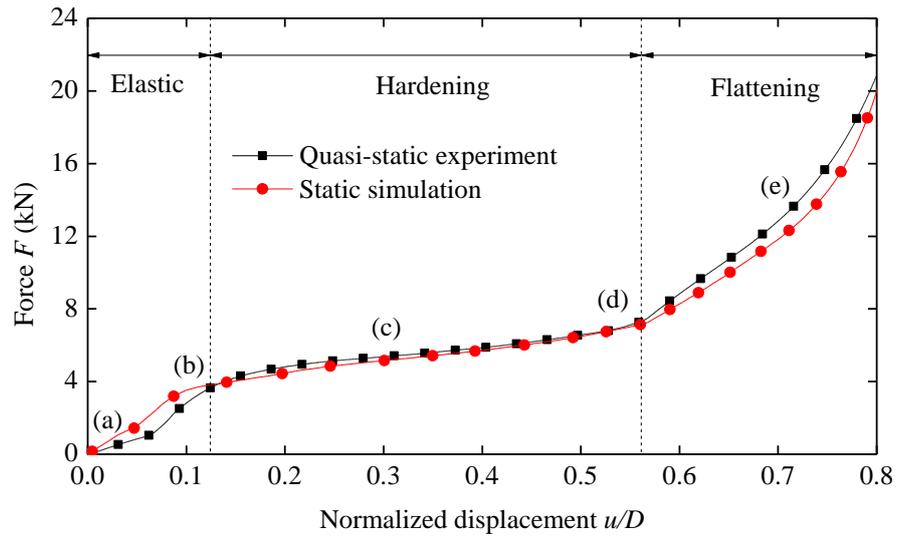

**Figure 11** Force response of a dumbbell-shaped tube under compression, in which the deformed configurations obtained from simulation and experiment at points (a)-(e) correspond to Figs. 9(a)-(e) and 9(f)-(j), respectively.

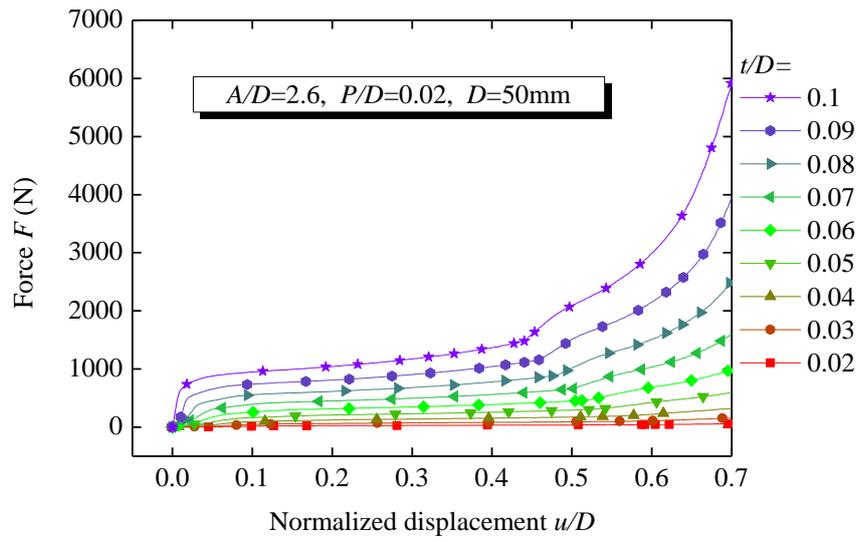

(a)

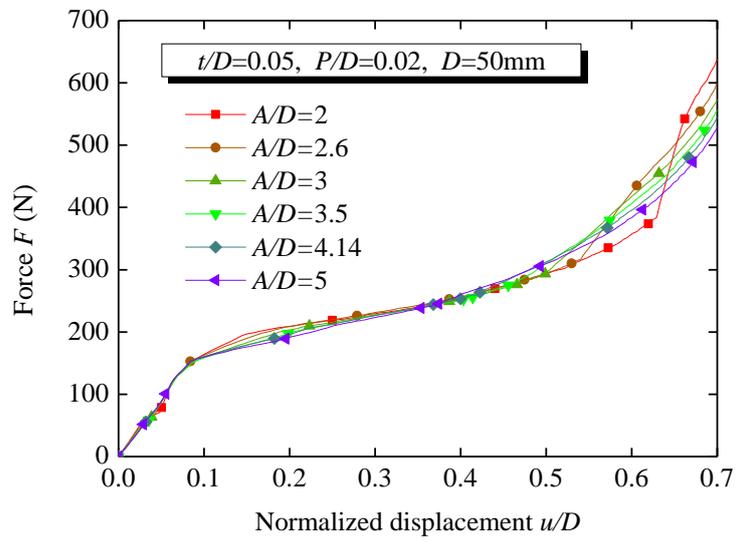

(b)

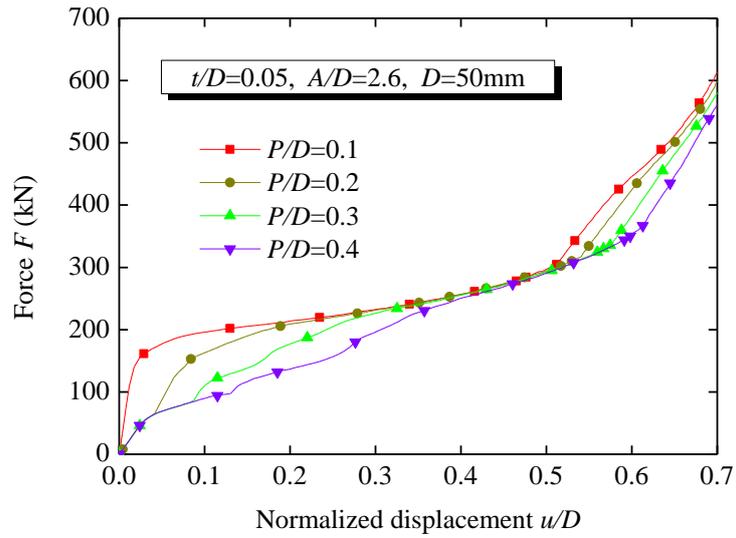

(c)

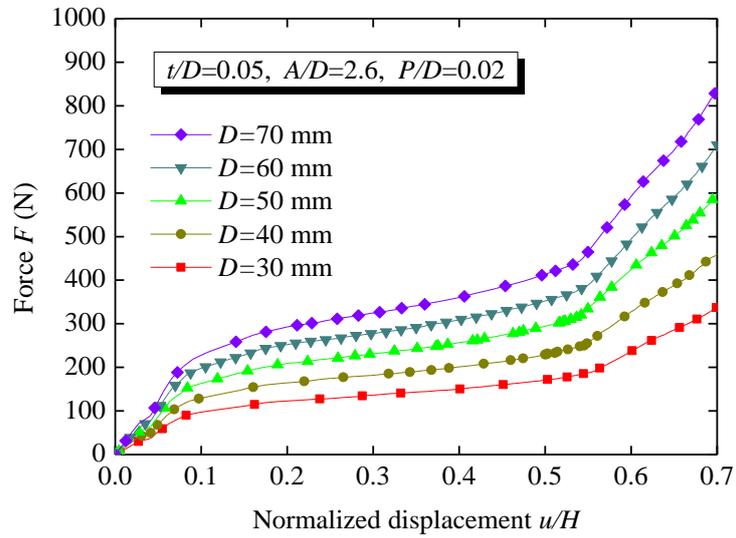

(d)

**Figure 12** Force responses of dumbbell-shaped tubes with different (a) thickness to diameter ratio *t*/*D*, (b) axis distance to diameter ratio *A*/*D*, (c) plate spacing to diameter ratio *P*/*D* and (d) diameter *D*.

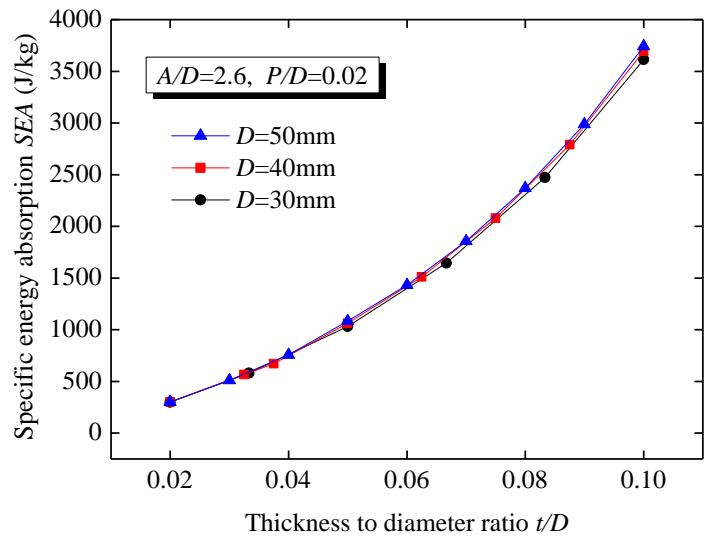

(a)

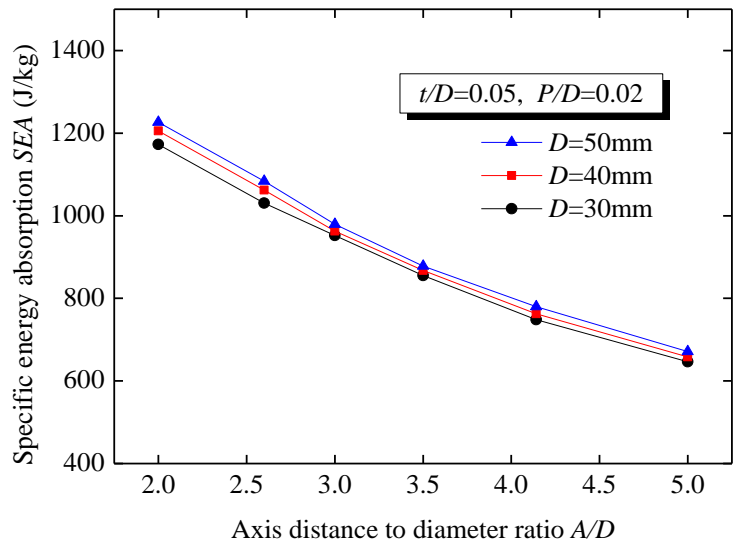

(b)

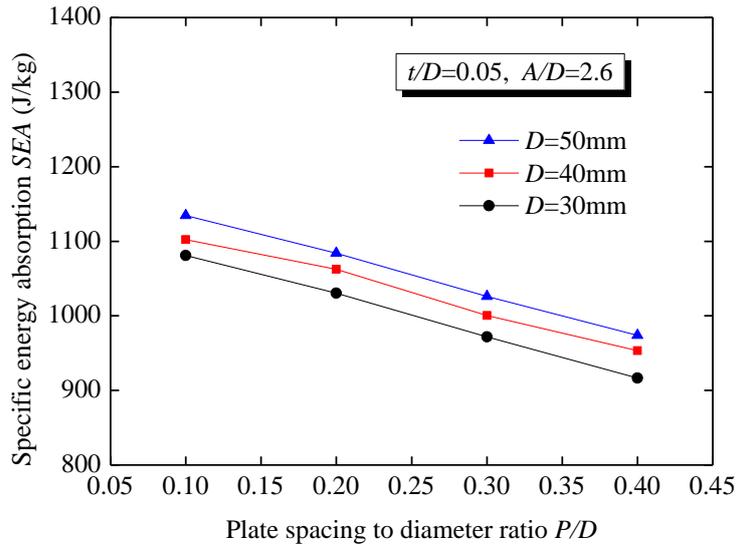

(c)

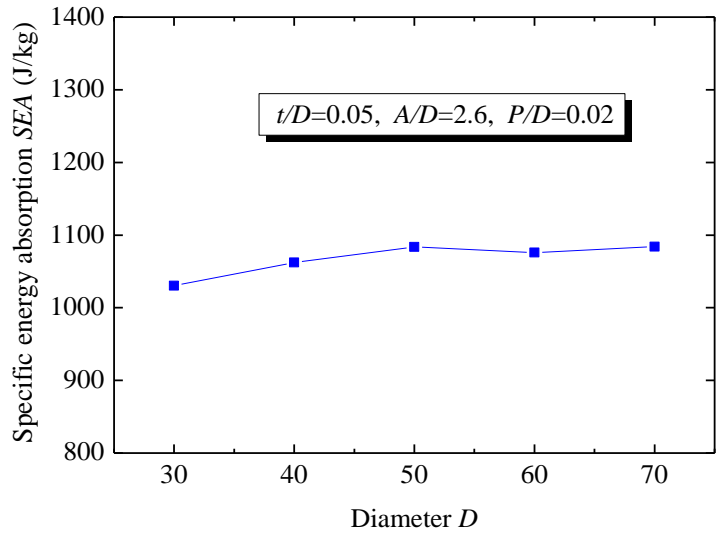

(d)

**Figure 13** Specific energy absorption of dumbbell-shaped tubes with different (a) thickness to diameter ratio *t*/*D*, (b) axis distance to diameter ratio *A*/*D*, (c) plate spacing to diameter ratio *P*/*D* and (d) diameter *D*.

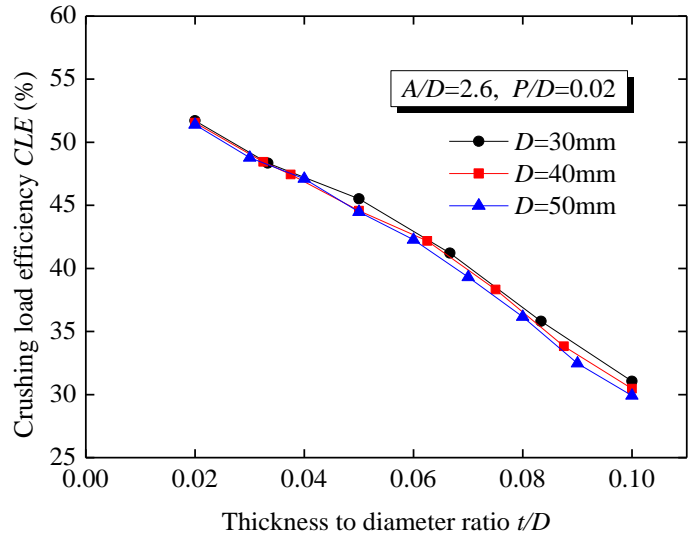

(a)

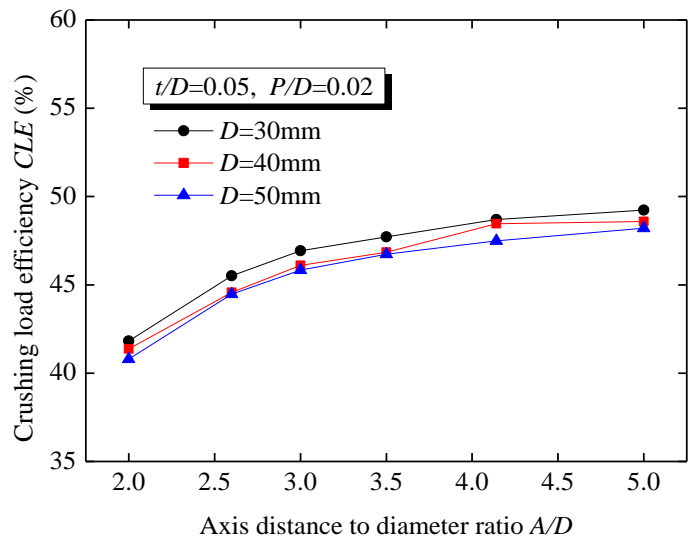

(b)

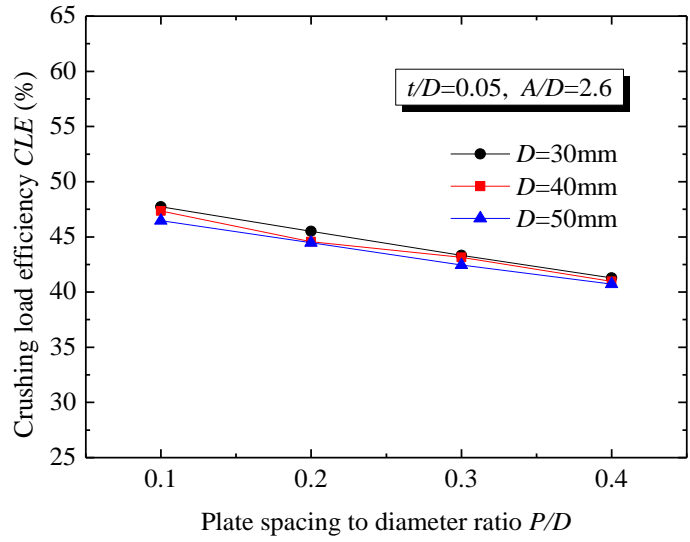

(c)

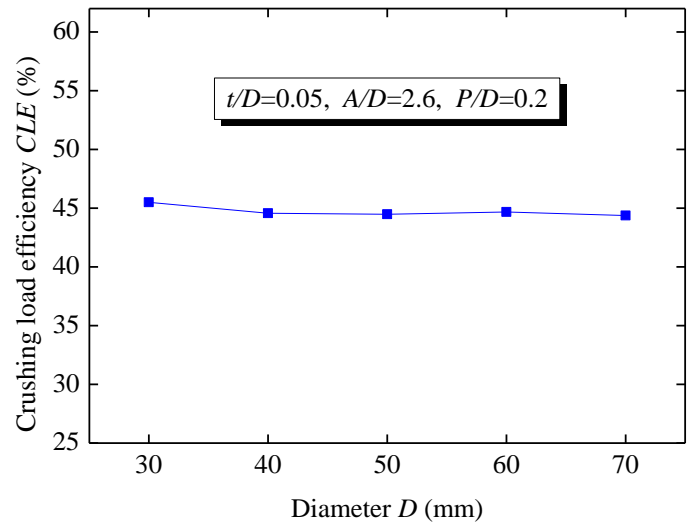

(d)

**Figure 14** Crushing load efficiency of dumbbell-shaped tubes with different (a) thickness to diameter ratio $t/D$, (b) axis distance to diameter ratio $A/D$, (c) plate spacing to diameter ratio $P/D$ and (d) diameter $D$.

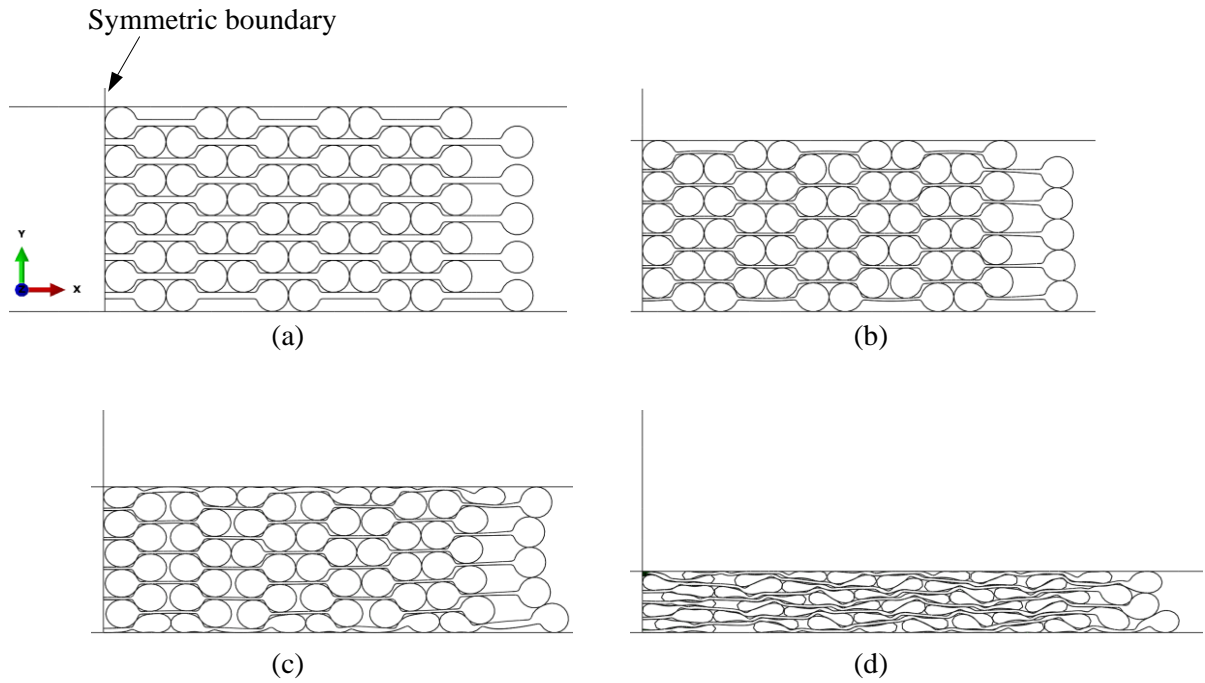

**Figure 15** Compressive simulation of a symmetric self-locked system with $n_{layer}=10$ and $n_{width}=7$ at normalized displacements of (a) $u/H=0$, (b) $u/H=0.17$, (c) $u/H=0.25$ and (d) $u/H=0.7$.

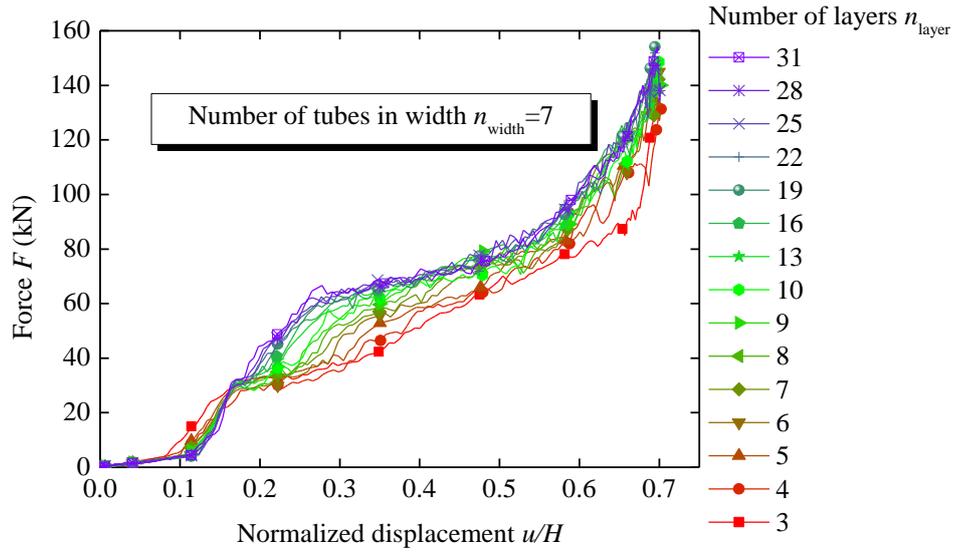

(a)

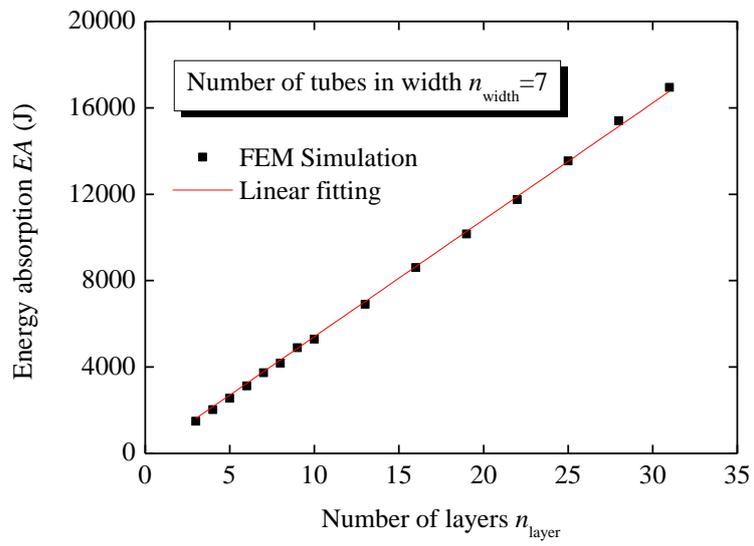

(b)

**Figure 16** Effect of number of layers on (a) force and (b) energy absorption.

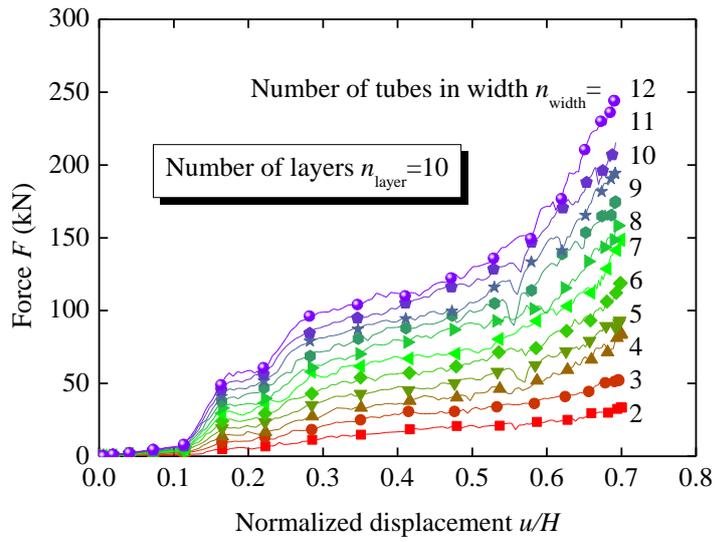

(a)

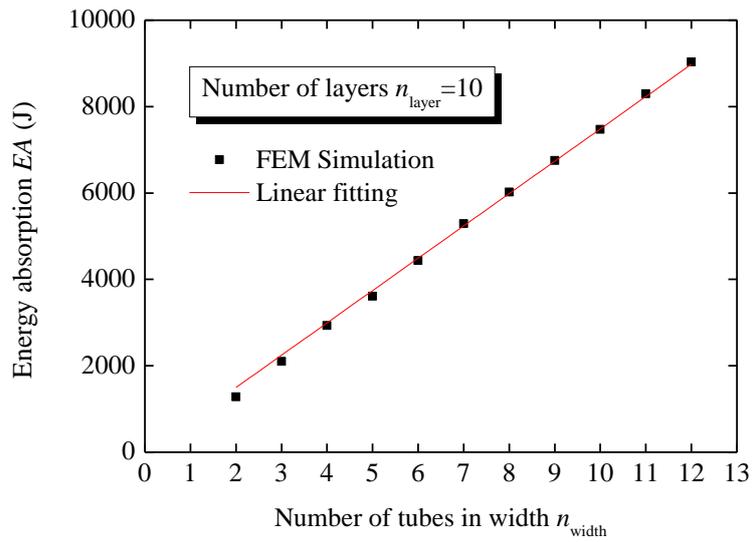

(b)

**Figure 17** Effect of number of tubes in width on (a) force and (b) energy absorption.

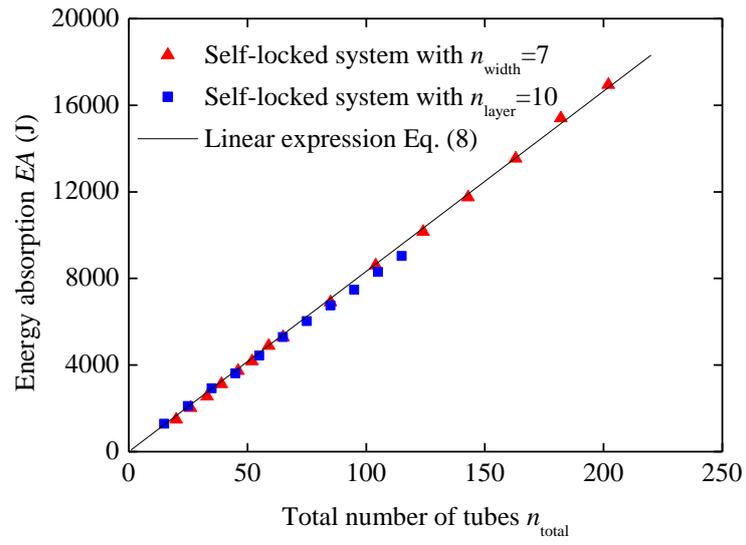

**Figure 18**   Energy absorption versus the total number of tubes.

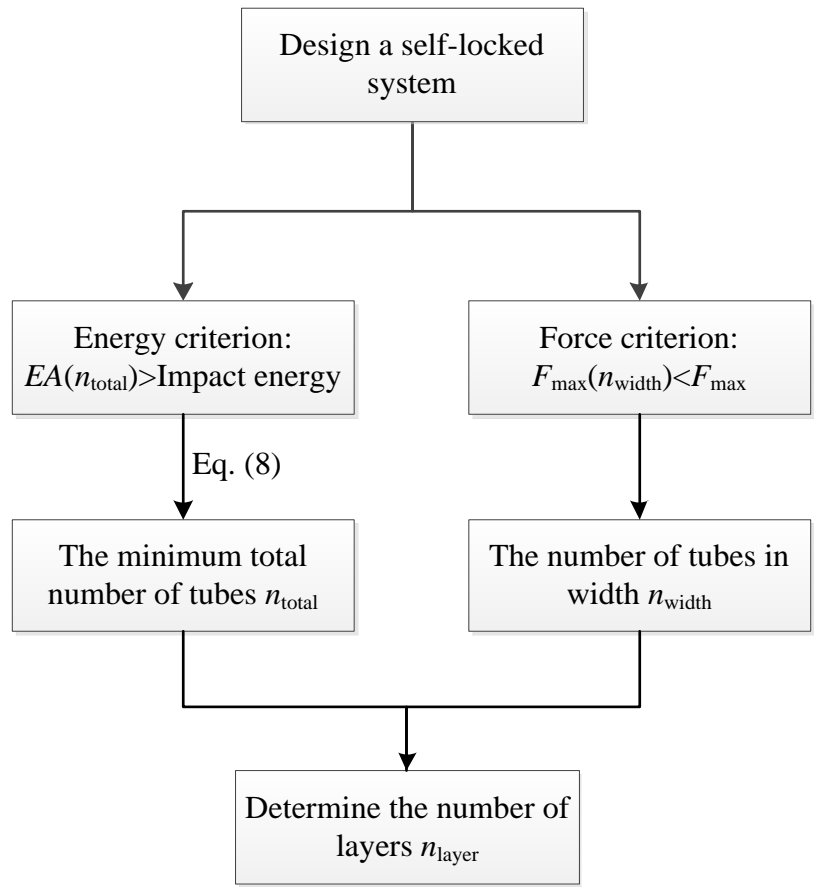

**Figure 19** A general guideline on the design of a rectangularly-stacked self-locked system.